\documentclass[useAMS,usenatbib]{mn2e}

\usepackage{graphicx,longtable,lscape}
\usepackage{rotating}
\title[PMS stars in Sh~2-284]{Pre-Main Sequence stars in the 
star forming complex Sh~2-284\thanks{Based on observations carried 
out at the European Southern Observatory (Paranal, Chile) under 
programme No. 074.C-0111. and partly at the Anglo-Australian Telescope in program
 07B/040 and 08B/003.}}
\author[F. Cusano et al.]{
F. Cusano$^{1}$\thanks{E-mail:fcusano@na.astro.it},
V. Ripepi$^{1}$, 
J. M. Alcal\'{a}$^{1}$,
D. Gandolfi$^{2}$, 
M. Marconi$^{1}$, 
S. Degl'Innocenti$^{3,4}$, 
\newauthor
~F. Palla$^{5}$,
E. W. Guenther$^{6}$ ,
S. Bernabei$^{7}$, 
E. Covino$^{1}$, 
C. Neiner$^{8}$, 
E. Puga$^{9,10}$, 
S. Hony$^{11}$ \\
$^{1}$ INAF-Osservatorio Astronomico di Capodimonte, Salita Moiariello16, 80131 - Napoli, Italy\\
$^{2}$ Research and Scientific Support Department, ESTEC/ESA, PO Box 299, 2200 AG Noordwijk, The Netherlands\\
$^{3}$ Dipartimento di Fisica, Universit\`a di Pisa, Largo B. Pontecorvo 3, 56126 Pisa, Italy \\
$^{4}$ INFN, Sezione di Pisa, Largo B. Pontecorvo 3, 56126 Pisa, Italy \\ 
$^{5}$ INAF-Osservatorio Astrofisico di Arcetri, Largo E. Fermi, 5, I-50125, Firenze, Italy \\
$^{6}$ Th\"uringer Landessternwarte Tautenburg, Sternwarte 5, D - 07778, Tautenburg, Germany \\ 
$^{7}$ INAF-Osservatorio Astronomico di Bologna, Via Ranzani 1, 40127 Bologna, Italy \\
$^{8}$ GEPI, UMR 8111 du CNRS, 5 place Jules Janssen, 92195 Meudon Cedex, France \\
$^{9}$ Katholieke Universiteit Leuven, Celestijnenlaan 200D, 3001 Leuven, Belgium \\
$^{10}$ Centro de Astrobiolog{\'{i}}a (CSIC-INTA), Ctra. de Torrej\'on km-4,E28850, Torrej\'on de Ardoz, Madrid, Spain\\
$^{11}$ CEA Saclay, Bat.609 Orme des Merisiers, 91191 Gif-sur-Yvette, France \\
} 
\begin{document}

\date{Accepted . Received}

\pagerange{\pageref{firstpage}--\pageref{lastpage}} \pubyear{2010}

\maketitle

\label{firstpage}

\begin{abstract}
{Located at the galactic anticenter, Sh~2-284 is a H{\sc ii} region which harbors 
several young open clusters; Dolidze~25, a rare metal poor (Z$\sim$0.004) young 
cluster, is one of these. Given its association with Sh~2-284, it is reasonable 
to assume the low metallicity for the whole H{\sc ii} region. Sh~2-284 is expected 
to host a significant population of Pre-Main Sequence (PMS) stars of both low and 
intermediate mass stars (Herbig Ae stars). We aim at characterizing these stars by 
means of a spectroscopic and photometric survey conducted with VIMOS@VLT and 
complemented with additional optical and infrared observations. In this survey we 
selected and characterized 23 PMS objects. We derived the effective 
temperature, the spectral energy distribution and luminosity of these objects; 
using theoretical PMS evolutionary tracks,  with the appropriate metallicity, we 
estimated the mass and the age of the studied objects.  We also estimated a distance of 4 Kpc for Sh 2-284 
by using spectroscopic parallax of 3 OB stars.    
From the age determination we concluded that triggered star formation is in act 
in this region. Our results show that a significant fraction of the young stellar objects (YSOs) may have 
preserved their disk/envelopes, in contrast with what is found in other recent 
studies of low-metallicity star forming regions in the Galaxy. Finally, among the 
23 bona fide PMS stars, we identified  8 stars which are good candidates to 
pulsators of the $\delta$ Scuti type. }
\end{abstract}

 
\section{Introduction}

Dolidze~25 is a young \citep[age $\approx$ 6 Myr;][]{turbide93} open cluster
associated with  the H{\sc ii} region Sh~2-284 \citep{sharpless1959}
 located in the galactic anticenter. On the basis of high-resolution
spectroscopy of three OB stars, \citet{lennon90} estimated that Dolidze~25
is deficient in metals by a factor of about 6 with respect to the Sun.
This result was confirmed later by \citet{fitzsimmons1992}. Dolidze~25 hence
represents a very rare case of galactic low-metallicity young cluster.
The association of Dolidze~25 to Sh~2-284, leads to the conclusion that
the H{\sc ii} region and the star forming sites in Sh~2-284 are characterized
by a low metallicity. \citet{delgado2009} proposed the presence of
two populations of stars in Dolidze~25. One composed by young stars of
3-4 Myr and the other one by older stars with an average age of 40 Myr.
The average cluster reddening derived by  \citet{delgado2009} is E(B-V)=0.78,
which is in good agreement with the value E(B-V)=$0.81\pm0.11$ estimated
by \citet{moffat1975}.

 It is well known that a great part of YSOs
\citep[ClassI and ClassII objects; see][]{Lad87} show infrared excess, which 
is interpreted in terms of the presence of a circumstellar disk/envelope. 
Infrared colour-colour (CC) and colour-magnitude (CM) diagrams are good 
diagnostic tools for the investigation of circumstellar matter around 
YSOs \citep[][and references therein]{Hart05, Lad06, Har07, Alc08}. 
Recently, \citet[][hereafter P09 ]{puga2009} carried out an infrared photometric survey of 
the star forming region in Sh~2-284, using the Infrared Array Camera 
\citep[IRAC;][]{fazio04} on board  the Spitzer Space Telescope \citep{werner2004}. 
Based on the [3.6]-[4.5] vs. [5.8]-[8.0] diagram P09 selected 338 
candidates to infrared-excess YSOs. They classified 155 and 183 as 
Class~I and Class~II IR sources, respectively and studied their spatial 
distribution. They found that the YSO candidates concentrate in several 
aggregates, one of which is projected on the field of Dolidze~25. 
They conclude that triggered star formation may be at work in the region.
Both the works by P09 and Delgado et al., do provide a photometric 
characterization of the sources, but no spectroscopic information on the
low and intermediate mass YSOs in Sh~2-284 and members of Dolidze~25 is 
available.
 
In this paper we present the results of a photometric and spectroscopic 
follow-up of Sh~2-284 using  VIMOS 
\citep[VIsual $\&$ Multi-Object Spectrograph,][]{lefevre2003} at VLT.
We aim at characterizing low and intermediate mass PMS stars of the 
region to study star formation in such low-metallicity environment.
Our interest is also focused on the identification of intermediate-mass 
(1.5M$_{\odot}$$\le$M$\le$4M$_{\odot}$) PMS stars falling in the 
$\delta$-Scuti instability strip \citep{breger72}. Since about ten years 
it is known that such stars, during their contraction towards the main 
sequence, cross the instability strip for more evolved $\delta$-Scuti 
stars \citep{breger98}. These PMS stars are thus subjected to stellar 
pulsation \citep[see][]{marconi98}. Therefore, their asteroseismological 
properties allow to put constraints on the PMS physical parameters such 
as luminosity, mass and effective temperature, as well as on the
internal structure, which is different from that of more evolved 
stars of the same mass.
 Good examples of known PMS $\delta$ Scuti
stars are the spectroscopic binary system RS Chamaeleontis \citep{bohm2009}, V351 Ori  \citep{ripepi2003} 
and IP Per \citep{ripepi2006}. 
However, the asteroseismological study of some 
of the $\delta$ Scuti PMS found here is still in progress and is
 deferred to another paper (Ripepi et al. 2010, in preparation).
 
The paper is organized as follows. In Sect.~2 we present our observations
and describe the data reduction. In Sect.~3 we discuss the results: 
the determination of physical parameters together with the selection 
of bona-fide PMS and their infrared properties. In the same section we give 
an estimate of the distance of the Sh 2-284 region by using spectroscopic parallax performed on  early-type stars. Finally, discussions 
and conclusions are presented in the Sect.~4 and 5.
 
\section{Observations and data reduction}
 
\begin{figure*}
\includegraphics[width=75.mm,height=170.mm,angle=270]{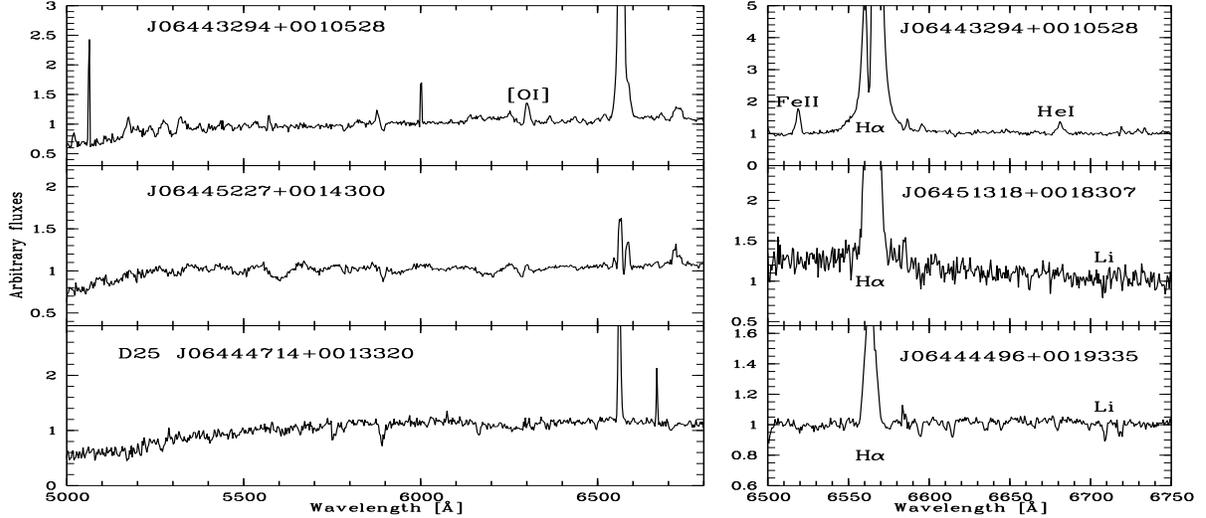}
 \caption{The MR spectra of three PMS stars in Sh~2-284
  are shown in the left panels. The right panels show the
  HR spectra of other PMS stars in the region. The later 
  cover the spectral region between 6500 and 6750~\AA, where 
  the H$\alpha$ emission and the lithium absorption lines can 
  be seen.
  Emission lines due to stellar wind ([OI]) and accretion
  process (H$\alpha$) are well detected in the spectrum of 
  J06443294+0010528.}
\label{fig:spec}
\end{figure*}
 
Both photometry and spectroscopy were obtained in service mode with
VIMOS at the UT3 of the VLT complex at Paranal (Chile). 
In the field of Sh~2-284 we obtained photometric data on the 7th October 2005 
in the $R$ and $I$ filters, for a total of 16 images. 
The  spectroscopic data were acquired the 4th February 2006 and consist 
of about 1500 spectra at low and medium resolution. In the following 
subsections we describe the details of the observations and the data 
reduction.
 
\label{sec:Observations}
 
\subsection{Photometry}
The VIMOS imager is composed by a mosaic of 4 CCDs with gaps of 
about 2~arcmin  for a total field of view (FOV) of 
$18^{'} \times 16^{'}$.  In order to image approximately 
$30'\times 30'$, which cover most of the central part of Sh~2-284 H{\sc ii} 
region, we performed 2 pairs of $R$ and $I$ frames with pointings 
shifted by about 6 arcmin between one pair and the other. Equal 
exposures of 60\,s were performed in both filters.  

The images were processed following standard techniques (bias
subtraction and flat-field correction) using the IRAF\footnote{IRAF 
is distributed by the National Optical Astronomical Observatories,
which are operated by the Association of Universities for Research
in Astronomy, Inc., under cooperative agreement with the National
Science Foundation} package. The fringing in the $I$-band images
was corrected by adopting a well-suited fringing map kindly 
provided to us by E. Held and S. Berta and following the recipe 
by \citet{alcala02}. The images were then astrometrically calibrated 
using the standard IRAF procedures and the USNO-A2.0 as reference 
catalogue. The resulting astrometric accuracy is of the order of 
0.35\,arc-sec RMS. 

PSF photometry was performed with the DaophotII/Allstar 
packages \citep{stetson87,stetson92}.  The photometric calibration was 
executed by using the standard fields L98 and L110 \citep{landolt92}. 
For this purpose we used the extended list of standards provided by 
P.B. Stetson\footnote{http://cadcwww.dao.nrc.ca/standards/}, 
which are sufficiently large in number of stars to include a 
significant number of objects ($>$100) in each CCD, allowing 
the determination of a colour equation independently for all the chips. 
As a result of this procedure, we obtained four linear and four 
quadratic calibration equations in the $R$ and $I$ bands, respectively.  
The typical error of the transformations is of the order of 0.01-0.02\,mag.  
We then applied these transformations to the instrumental magnitudes 
of the objects extracted from our scientific images, and derived 
the magnitudes in the standard Jonhson-Cousins system. 

Our photometric catalog was complemented with measurements in 
other bands from literature. We used: 1) $BV$ photometry of the
same field provided by the COROT EXO-DAT group \citep{meunier2009}; 
2) the $JHK$ photometry from the 2-Micron All-Sky Survey (2MASS) 
point-source catalog \citep{cutri03}; 3) the mid-infrared Spitzer
photometry at 3.6, 4.5, 5.8, and 8 $\mu$m, obtained under the 
program 3340 (P09).  
As  result, we end up with a photometric catalogue with a large 
wavelength coverage, from the optical to the mid-infrared,
which is crucial to investigate the PMS nature of our sources in 
terms of infrared excess in their spectral energy distribution.

\subsection{Spectroscopy}
\label{spec_obs}
The multi-slit spectroscopy was acquired using the Medium
Resolution (MR), High Resolution (HR) Red and the HR
Orange\footnote{The Orange grism was used in quadrant 4 due to the
lack of a HR red grism for this CCD} grisms. The wavelength
coverage and resolution corresponding to these grisms are: 4800-10000
$\rm{\AA}$ with 2.5 $\rm{\AA}$/pix, 6500-8750 $\rm{\AA}$ with 0.6
$\rm{\AA}$/pix and 5150-7600 $\rm{\AA}$ with 0.6 $\rm{\AA}$/pix for
MR, HR red and HR orange grisms, respectively. Note, however, that 
the spectral coverage for each object depends on the position of 
the corresponding slit relative to the center of the VIMOS quadrants.  
In each CCD we were able to observe around 40 and 30 objects 
simultaneously with the MR and HR grisms, respectively. In total, 
we obtained 912 spectra with the MR red and  594 spectra with the 
HR Red and Orange grisms. The exposure time was 900~s for the MR 
grism with an average S/N $\sim$40-50, whereas for the HR grism the 
exposure time was 2000~s with an average S/N $\sim$20-30.

The bias subtraction, flat field correction, sky subtraction, wavelength 
calibration of the spectra were performed using the ESO VIMOS automatic 
pipeline. Subsequent extraction of the spectra was executed using the 
IRAF routine $apall$. As a test of the goodness of the data reduction 
performed using the pipeline, we also processed a few spectra by hand 
using the standard IRAF routines. 
The comparison of the results showed that the automatic pipeline provides 
scientifically useful spectra. Unfortunately, the spectral region with 
$\lambda > 7000 \rm{\AA}$ of the low-resolution spectra is useless due 
to strong fringing. Some examples of spectra are shown in Fig.~\ref{fig:spec}. 

Finally, we calibrated all the spectra in relative flux. To this 
aim flux standard stars were observed during the same nights as 
the scientific frames. The standard stars were reduced using IRAF 
standard routines. A response function was derived for each of 
the four VIMOS quadrants. 

Due to constraints on the software used to prepare the VIMOS masks 
(VIMOS Mask Preparation Software, VMMPS), it was not possible to configure 
slits on many interesting candidates selected from optical colour-magnitude 
diagrams, preventing the observation of 
high-priority targets. However, it was possible to automatically assign 
slits to other objects in the field of the cluster.

 Additional low-resolution spectroscopy of three OB stars  was performed in January 2008 with the AAOmega multi-object facility mounted 
at the AAT-3.9\,m telescope of the Anglo-Australian Observatory. Two out of these three stars were investigated by \citet{lennon90} and are indicated in this paper 
as   star number 15 and 22.
These observations are part of a large spectroscopic follow-up program devoted to
 the study of the stellar population in the  $CoRoT$ fields and will be described in detail in a forthcoming paper (Gandolfi et al., in preparation).
  The \emph{580V} and \emph{385R} low-resolution gratings were used for the blue and red arm of the spectrograph, respectively, yielding a spectral coverage of 
  about 3700~{\AA} ($3730 - 7430$\,\AA~) with a mean resolving power of $R \approx 1300$.

\section{Results}

The spectroscopy in this work represents a step forward with respect
to the previous photometric studies of Sh~2-284. Thus, we discuss first
the spectroscopic results and then use them to further investigate 
the photometric properties, from mid-IR to optical, of the YSO 
candidates in the region.

\subsection{Spectroscopically identified PMS stars}
\label{sec_sel_pms}
 
The main criteria used for the selection of PMS stars are the presence 
of H$\alpha$ emission plus near and mid IR excess of emission relative to 
normal stars of the same spectral type. Both processes trace the presence 
of circumstellar 
material around YSOs. The H$\alpha$ emission on one side, indicates 
accretion from the circumstellar disk towards the star. The infrared 
excess comes indeed from the thermal disk emission. 
To detect young stars, an additional criterion is the presence in the 
spectra of the Li{\sc i} ($\lambda 6708\rm{\AA}$) absorption line, for 
stars with spectral type later than about G5. The Lithium, in fact, 
is efficiently burned in the first million years of PMS stellar 
evolution. 

\subsubsection{H$\alpha$ emission}
First, an inspection of all the available spectra was performed searching for 
$\rm{H\alpha}$ emission-line stars. This inspection led to the identification 
of 35 emission-line objects, which are reported in Table~\ref{tab1} and \ref{tab2} together 
with the photometric measurements. The equivalent width of the H$\alpha$ 
line was then measured using IRAF. Such measurements, reported in 
Table~\ref{tab3} and \ref{tab4}, were performed by Gaussian fitting, defining the continuum 
by fitting a low order spline to selected spectral regions in both sides of 
the $\rm{H\alpha}$ line. The errors in such measurements come mainly from 
the uncertainty in the continuum definition when selecting the regions near 
the line. Since the H$\alpha$ criterion alone cannot discriminate between 
PMS stars and field emission-line stars, we used additional criteria 
based on the available IR photometry (see Section~\ref{col-mag-diags}).

\subsubsection{Spectral types}
\label{sec_spt}
The spectral typing was performed on the H$\alpha$ emitting-stars by using 
calibrations in the literature between equivalent width of some spectral 
lines versus spectral type as follows: 
 i) the blend between Ca I and Fe I at 5270\AA~ \citep{hernandez04} and 
ii) the Na I line at 5892 \AA~ \citep[only for spectral type later than K0][]{tripicchio97}.
As a check of the spectral type determinations we have also performed 
a $\chi^2$ minimization of a grid of spectral templates on  the 
available spectra of the PMS candidates. This procedure is similar to the
one described in \citet{frasca03}. The derived spectral types are reported 
in Tab.~\ref{tab3} and \ref{tab4} together with the equivalent width of the 
Ca{\sc i}+Fe{\sc i} blend and the Na{\sc i} line. The average uncertainty on 
spectral classification is estimated to be of one and two subclasses for high and 
low S/N spectra, respectively. The spectral type was used to derive the 
individual reddening and spectral energy distribution (SED) of the PMS 
stars (See Section~\ref{sec_SEDs}).
 
\subsubsection{Lithium absorption}
The presence of Li{\sc i} ($\lambda 6708\rm{\AA}$) absorption feature in the 
spectrum of late-type stars, can be used as an additional check for the PMS
nature of the stars. In the low-resolution spectra the line is blended with 
neutral iron lines. Thus, the line should be sought in the available VIMOS 
high-resolution spectra. Among the stars selected above, only 10 have 
high-resolution VIMOS spectra, but only five have sufficient S/N to detect 
the Li{\sc i} line. We unambiguously detect the line in all of them.  
Examples are shown in Fig.~\ref{fig:spec},
 adding support to our selection criteria.

\subsubsection{Vetted PMS stars}
On the basis of the above arguments 23 bona-fide PMS objects were identified. 
One of these, namely J06450075+0013356, was rejected as PMS star by 
\citet{delgado2009}, but our medium resolution spectrum (R$\sim2500$) shows 
strong H$\alpha$ emission and lithium absortpion (see Table \ref{tab3}). 
In addition, its SED shows infrared excess (See section~\ref{sec_SEDs} 
and Figure \ref{img:sed}). The 23 PMS stars are listed in Table~\ref{tab3}.  
 
Other two objects that show H$\alpha$ emission, namely J06453567+0018473 
and J06451118+0011402, match fairly well the age of PMS stars in 
the field of Dolidze~25 when assuming the distance of 4.0\,Kpc 
(see Section~3.1.5). However, since we lack their medium 
resolution spectrum, we cannot confirm or reject their PMS nature based 
on the Li absorption line. We thus refer to these two objects as potential 
PMS stars.
 
The remaining 10 H$\alpha$ emission line objects correspond to active 
field K and M-type stars, unrelated to the star forming region, but 
we also provide their coordinates and photometry in Table~\ref{tab2} and the spectral types in Table~\ref{tab4} . 
 
11 of the 23 PMS stars coincide with YSO candidates by P09,
which means that their selection missed about half of the PMS stars.
However, in order to be able to apply the selection criterion adopted 
by P09, it is necessary that the sources are detected 
in all four IRAC bands. Note that 6 of the PMS stars were not detected 
in at least one IRAC band, while other 6 have mid-IR colours typical 
of Class~III sources. The latter is further discussed in 
Section~\ref{col-mag-diags}.

\begin{figure*}
\begin{center}
\begin{tabular}{cc}
\multicolumn{1}{l}{\includegraphics*[height =5.7cm]{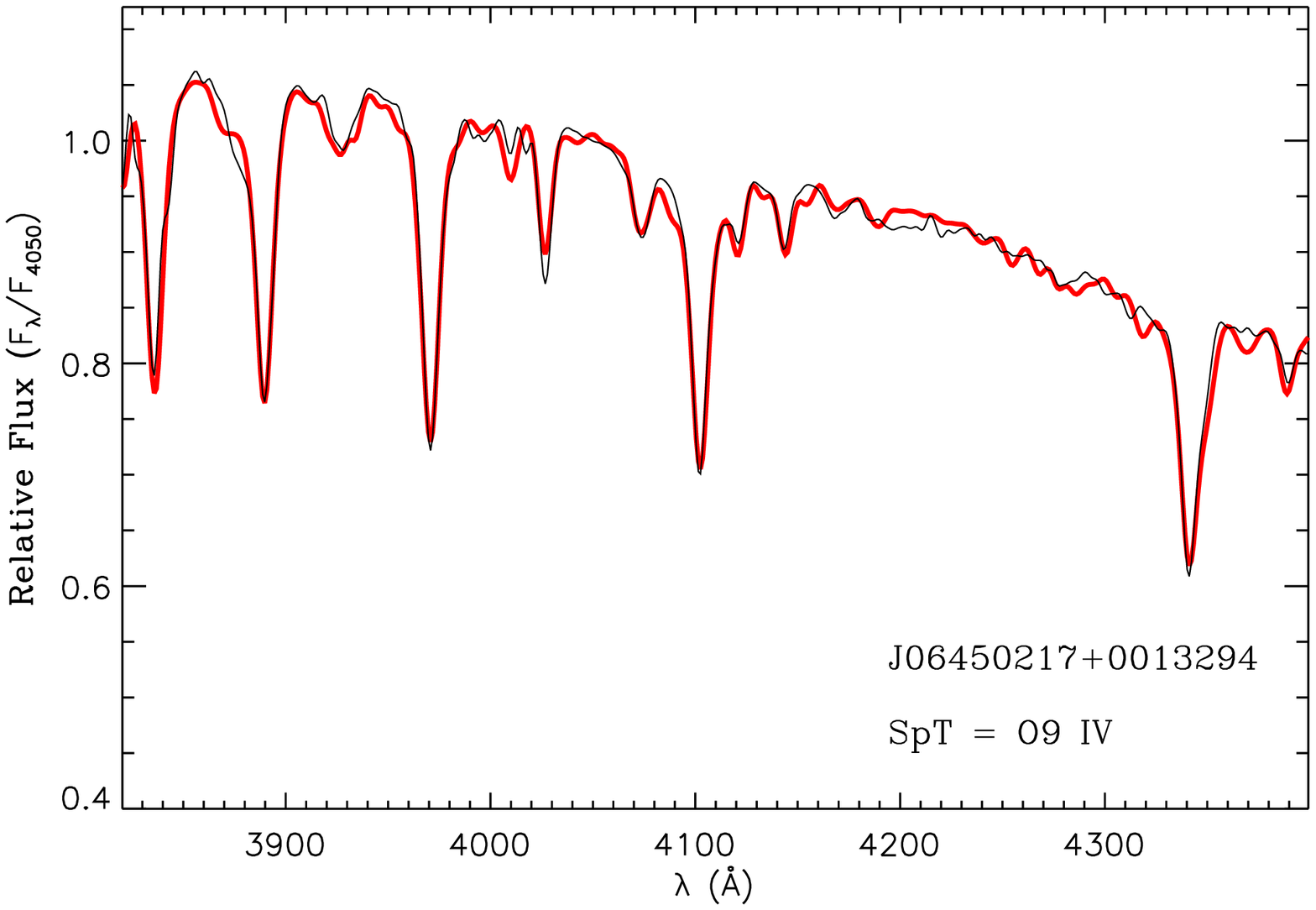}}&\multicolumn{1}{l}{\includegraphics*[height =5.7cm]{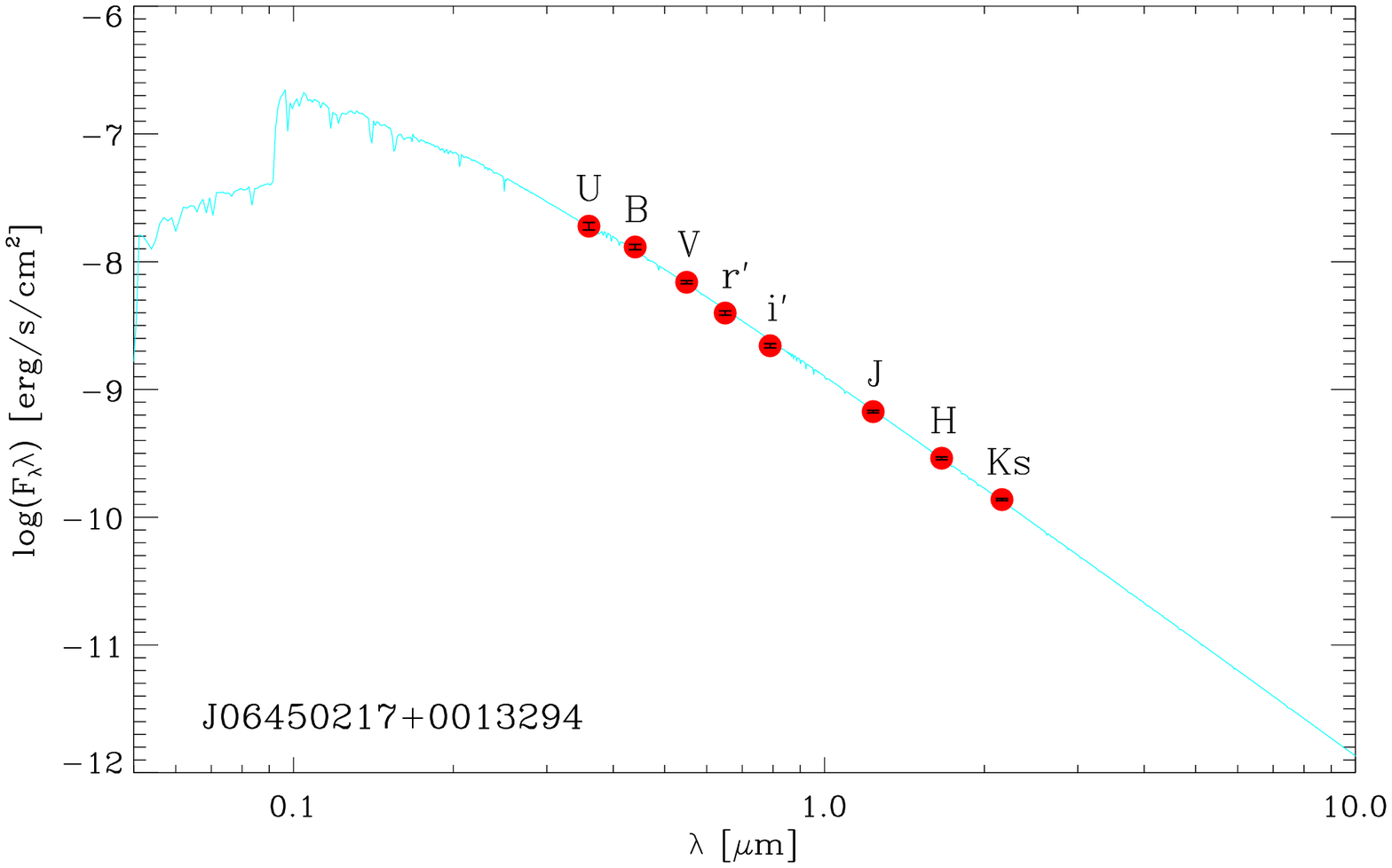}}\\
\end{tabular}
\end{center}
\caption{Left: AAOmega spectrum of J06450217+0013294 (thin black line). The best fitting O9 IV template is over-plotted with a thick red line. The spectra were  normalised to the
flux at 4050 \AA. Righ: SED of  J06450217+0013294. The red filled dots mark the dereddened optical and near-infrared fluxes. The model spectum from \citet{kurucz1979}, with the same effective temperature as the star, is over-plotted with a light blue thin line.}
\label{img:template}
\end{figure*}

\subsubsection{On the distance of Sh~2-284}
 There has been controversy on the heliocentric distance determinations 
of Sh~2-284. \citet{lennon90} and \citet{turbide93} give an estimate of $\sim$6~Kpc.   
\citet{russeil2007} determined a distance of $6.03\pm1.16$Kpc. They claim that 
Sh~2-284 belongs to a broader complex at $7.89\pm0.27$Kpc, which also includes 
the H{\sc ii} regions Sh~2-283, 285 and 286 \citep{sharpless1959}.  A spectrophotometric distance of $5.2\pm0.8$Kpc 
for Sh~2-284 was also determined by \citet{avedisova1984}. However, all these studies
did not account for the low-metallicity. More recently, \citet{delgado2009} 
estimated a distance of 3.6~Kpc by using low-metallicity isochronal fitting 
on $UBV$ CM diagrams of Dolidze~25, but they claim that 
the distance could be as high as 4~Kpc when considering sistematic effects
on metallicity determinations. In order to further investigate the distance of the Sh~2-284 region, we
performed a detailed analysis of three early-type stars in the field of 
Dolidze~25. These stars are 2MASSJ06450217+0013294, 2MASSJ06451043+0011164,
and 2MASSJ06441748+0020315. The first two stars are  also identified as star number 15 and 22  by 
\citet{moffat1975}, respectively. The spectral type and luminosity class of 
the target stars were derived by fitting the AAOmega spectra with a suitable 
grid of observed stellar templates (see example Figure \ref{img:template}), as described in \citet{frasca03} and 
\citet{gandolfi2008}. The effective temperature (T$_{\rm{eff}}$) and absolute 
magnitude (M$_{\rm V}$) of each star was  assigned using tabulated scales 
from the compilation of \citet{straizys1981}. A good agreement was found 
between the adopted M$_{\rm V}$ values and the absolute  magnitudes derived 
using the empirical H$_{\gamma}$-M$_{\rm V}$ calibration from \citet{millard1985}. 
We constructed the spectral energy distribution (SED) of the target 
stars (Figure \ref{img:template}) by merging the EXO-DAT $BVr^\prime i^\prime$ broad-band photometry 
\citep{meunier2009} with the U-magnitudes from \citet{moffat1975} and the 
$JHKs$ near-infrared photometry from the 2MASS catalogue. Following the
two-parameter method described in \citet{gandolfi2008}, we derived the 
interstellar extinction (A$_{\rm V}$) and the total-to-selective extinction 
R$_{\rm V}$=A$_{\rm V}$/E$_{\rm {B-V}}$ by fitting simultaneously the 
observed colors encompassed by the SED with synthetic colors computed 
using Kurucz 1979's atmosphere models with the same spectral type and 
luminosity class as the stars. The low-metallicity is taken into account 
by applying a correction to the absolute magnitude, appropriate 
for the metallicity of the region. Such correction, on the order of 0.6\,mag, 
was determined using the relationship by \citet{vandenberg1989} and adopting 
[Fe/H]=-0.55. We checked that this correction is valid for the range of 
intrinsic colors of the three stars by using the isochronal tools described
by \citet{marigo2008}. Finally, the individual distance to the stars was 
estimated using the metallicity-corrected M$_{\rm V}$ and A$_{\rm V}$ values, 
as well as the observed EXO-DAT V-magnitude. The derived physical parameters 
of the target stars are listed in Table~\ref{Results}. The estimated average distance of 4~Kpc, 
is in agreement with the upper value claimed by \citet{delgado2009}. 
Such distance places Dolidze~25 at a galactocentric distance of 12~Kpc. These values 
of galactocentric distance and metallicity are also consistent with the 
relationship between metallicity versus galactocentric distance for open 
clusters observed by \citet{sestito2008}. Thus, we adopt a heliocentric 
distance of 4~Kpc for subsequent analysis in this paper. 
\begin{table*}
  \caption{Physical parameters of the stars used to determine the distance of Sh~2-284. The fifth column reports the observed V-magnitude
   taken from the EXO-DAT \citep{meunier2009}.}
 \begin{center}
  \begin{tabular}{llccccccccc}
  \hline
  \hline
  \noalign{\medskip}
   id (2MASS)  &      id$^1$ &   $\alpha$ (J2000)& $\delta$ (J2000)      &   V      &     SpT      & T$_{\rm {eff}}$  &  M$_{\rm V}$  &    A$_{\rm V}$  &  R$_{\rm V}$  &   d$^{*}$          \\
	       &             &                   &                       &   [mag]     &              &     [K]    &        [mag]      &     [mag]           &                   &   [pc]         \\
  \noalign{\medskip}
  \hline
  \hline
  \noalign{\smallskip}
 J06450217+0013294     & 15         &     06 45 02.18  &	00 13 29.5   &   11.66       &    O9\,IV   &  33\,000   &      $-4.9\pm0.2$ &   $2.92\pm0.03$     &  $3.30\pm0.05$    &   $4060\pm500$  \\
  J06451043+0011164    &  22        &     06 45 10.44 & 	00 11 16.4   &  12.07	        &    B1\,IV   &  25\,000   &      $-3.9\pm0.2$ &   $2.27\pm0.08$     &  $3.20\pm0.20$    &   $4172\pm500$ \\
J06441748+0020315     &            &      06 44 17.49   &	00 20 31.5   &  11.93      &B0e\,IV    &    31\,000            &  $-4.4\pm0.2$    & $2.60\pm0.10$       &  $3.60\pm0.20$    & $4197\pm500$  \\

  \noalign{\smallskip} 
  \hline
  \hline
  \end{tabular}
  \end{center}
\begin{flushleft}
$^1$ from \citet{moffat1975}                                              \\
  $^{*}$: distance estimated by adding 0.6\,mag to M$_{\rm V}$ as a correction due to low-metallicity. 
 \end{flushleft}
  \label{Results}
\end{table*}

\subsection{Photometric properties}
\label{col-mag-diags}
 
\subsubsection{Mid-IR colour-magnitude and colour-colour diagrams}
 
\begin{figure*}
\begin{center}
\begin{tabular}{cc}
\multicolumn{1}{l}{\includegraphics*[height =8.7cm]{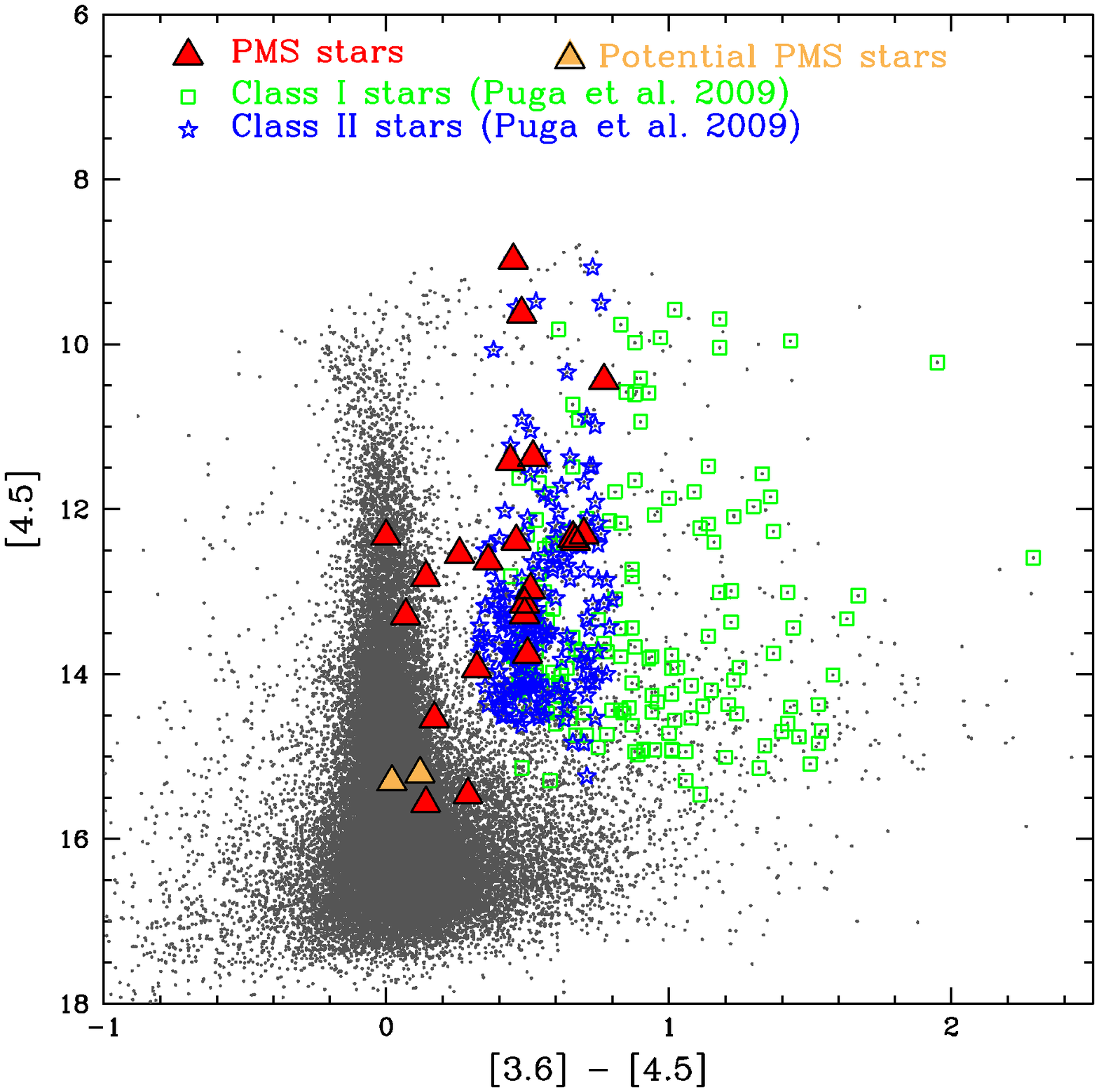}}&\multicolumn{1}{l}{\includegraphics*[height =8.7cm]{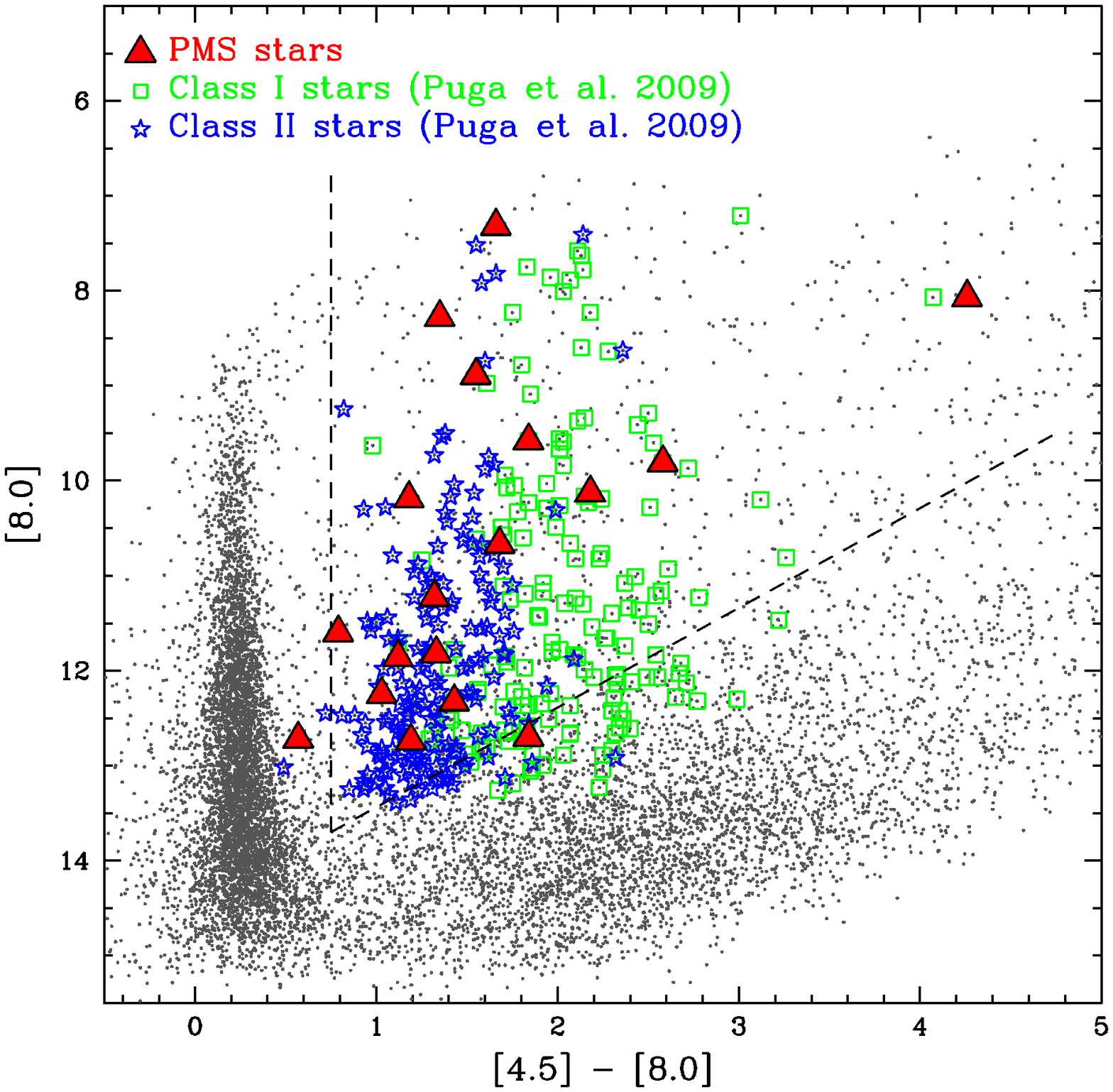}}\\
\multicolumn{1}{l}{\includegraphics*[height =8.7cm]{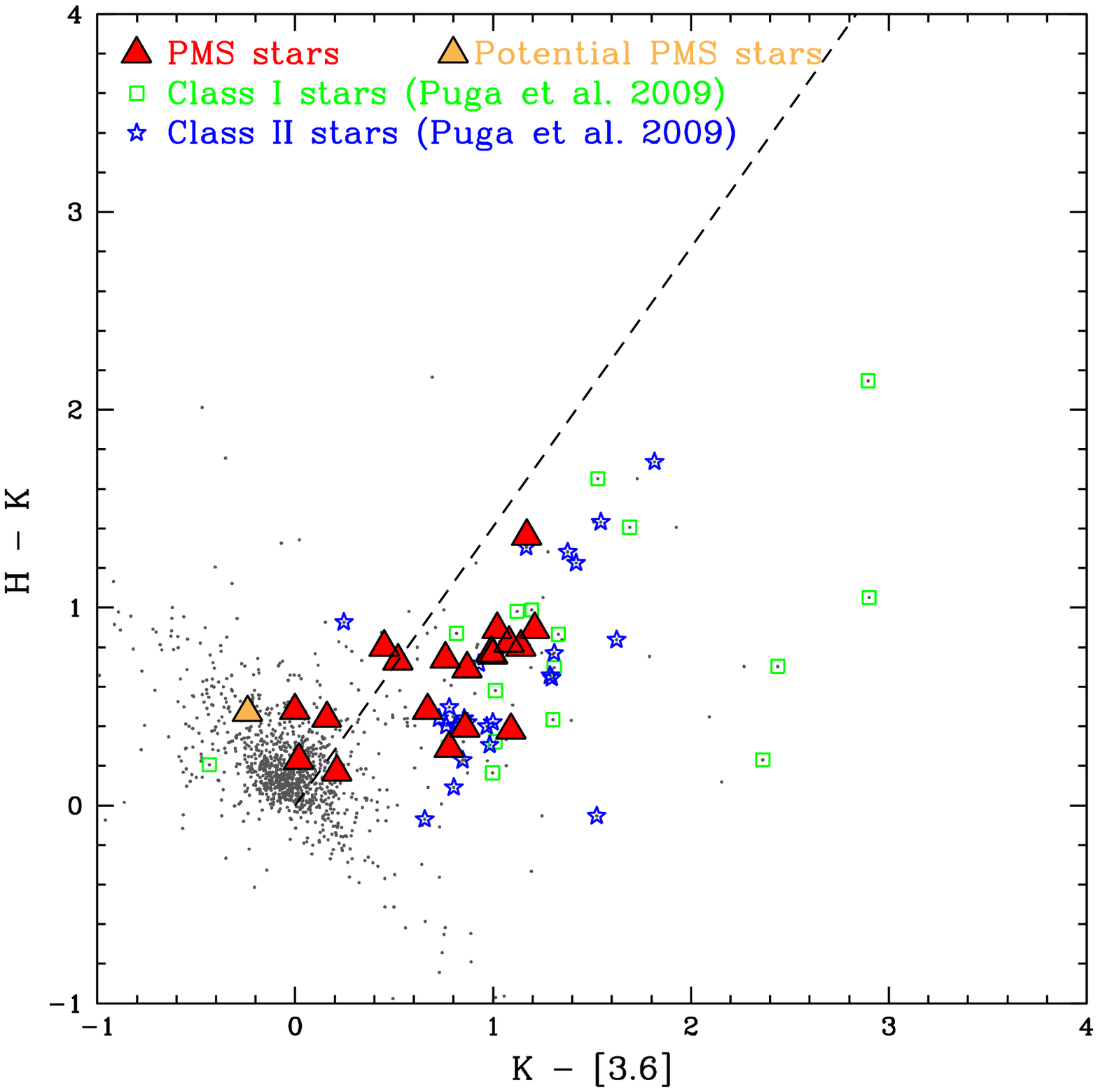}}&\multicolumn{1}{l}{\includegraphics*[height =8.7cm]{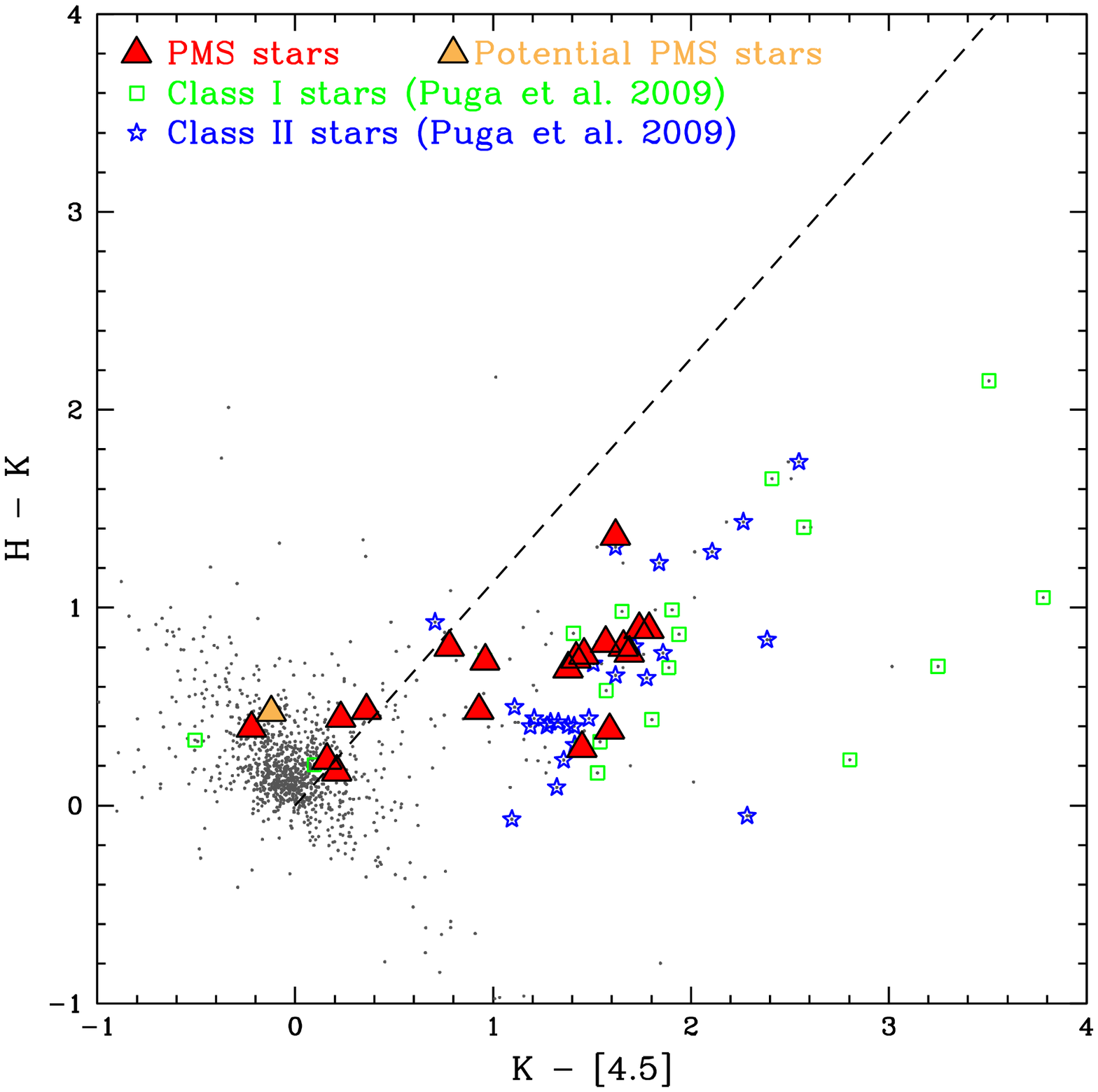}}\\
 \end{tabular}
\end{center}
\caption{IR colour-magnitude and colour-colour diagrams of the sources detected 
 with Spitzer in Sh~2-284 (gray dots). The Class~I and Class~II YSOs selected 
 by P09 are colour-coded as in the labels. The PMS stars 
 spectroscopically detected in this work (See Section~3.1) are 
 overplotted as red triangles. The criterion adopted by the c2d survey 
 \citep{eva03,eva09} for the selection of candidates to YSOs, based on 
 IRAC-only data \citep[e.g.][]{por07}, is represented with the dashed lines 
 in the [8.0] vs. [4.5]-[8.0] CM diagram. The slope of the dashed line in 
 the 2MASS-Spitzer CC diagrams corresponds to the colour-excess ratio by 
 \citet{Fla07}.}
\label{img:diaspit}
\end{figure*}

The classification of the sources as Class~I and Class~II YSOs by 
P09 was based solely on the [3.6]-[4.5] vs. [5.8]-[8.0] 
diagram, but other combinations of colours and magnitudes can be used 
to check the selection and classification, by also incorporating the 
near-IR 2MASS data. 
The two upper panels of Fig.~\ref{img:diaspit} show the Spitzer CM diagrams 
of all the sources detected in Sh~2-284, while the lower panels show the 
near-mid IR CC diagrams of those sources with 2MASS data. Most of the 
YSO candidates selected by P09 satisfy the ``cores-to-disk'' 
\citep[c2d;][]{eva03,eva09} criterion for the selection of YSO candidates 
based on IRAC-only data \citep[e.g.][]{por07}, but some might be extragalactic 
contaminants, in particular several Class~I candidates that fall below 
the inclined dashed line in the [8.0] vs. [4.5]-[8.0] CM diagram. 
The Class~I and Class~II sources are well distinguished in the Spitzer 
CM diagrams, while most of the spectroscopically identified PMS stars 
have typical colours of Class~II and Class~III sources. In particular, 
six of the latter can be well distinguished in the [4.5] vs. [3.6]-[4.5] 
CM diagram, typically at [3.6]-[4.5] colours below 0.4\,mag. Note also
that the potential PMS stars have colours of Class~III sources. These two 
objects remained undetected at 8 micron.

Not all the the sources detected by Spitzer in the Sh2-284 region were also detected by 
the 2MASS. The objects matching the two catalogues are plotted in the CC diagrams shown in the low
panel of Fig.~\ref{img:diaspit}.
 The stars with normal photospheric colours, all basically 
at zero, are well distinguished from the YSOs, but as noted above several 
of the spectroscopically identified PMS stars posses normal photospheric 
colours. These objects represent part of the diskless PMS population of 
Sh~2-284.
 
\subsubsection{Near-IR colour-colour diagram}
 
\begin{figure*}
\begin{center}
\begin{tabular}{cc}
\multicolumn{1}{l}{\includegraphics*[height=8.7cm]{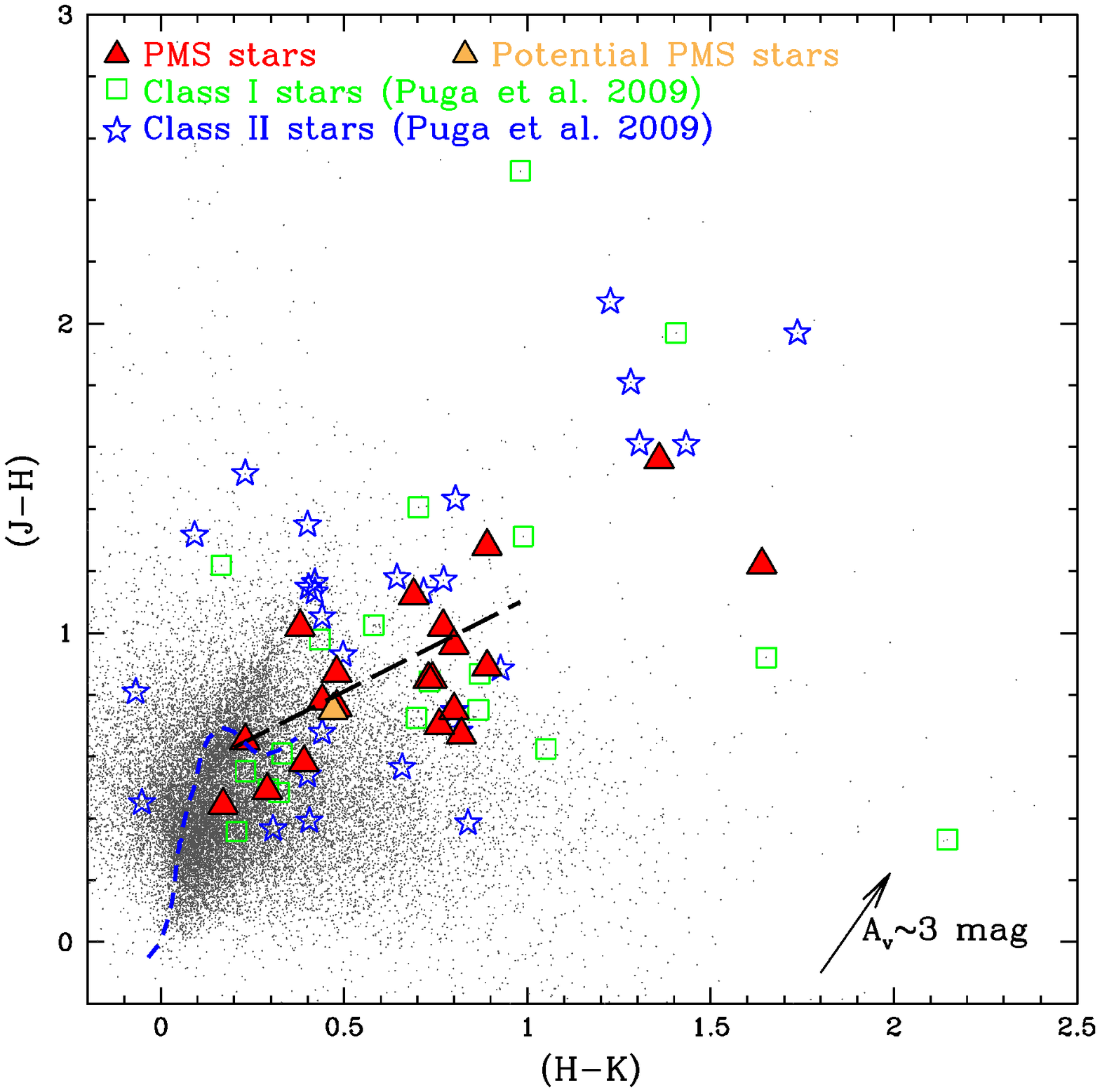}}&\multicolumn{1}{l}{\includegraphics*[height=8.7cm]{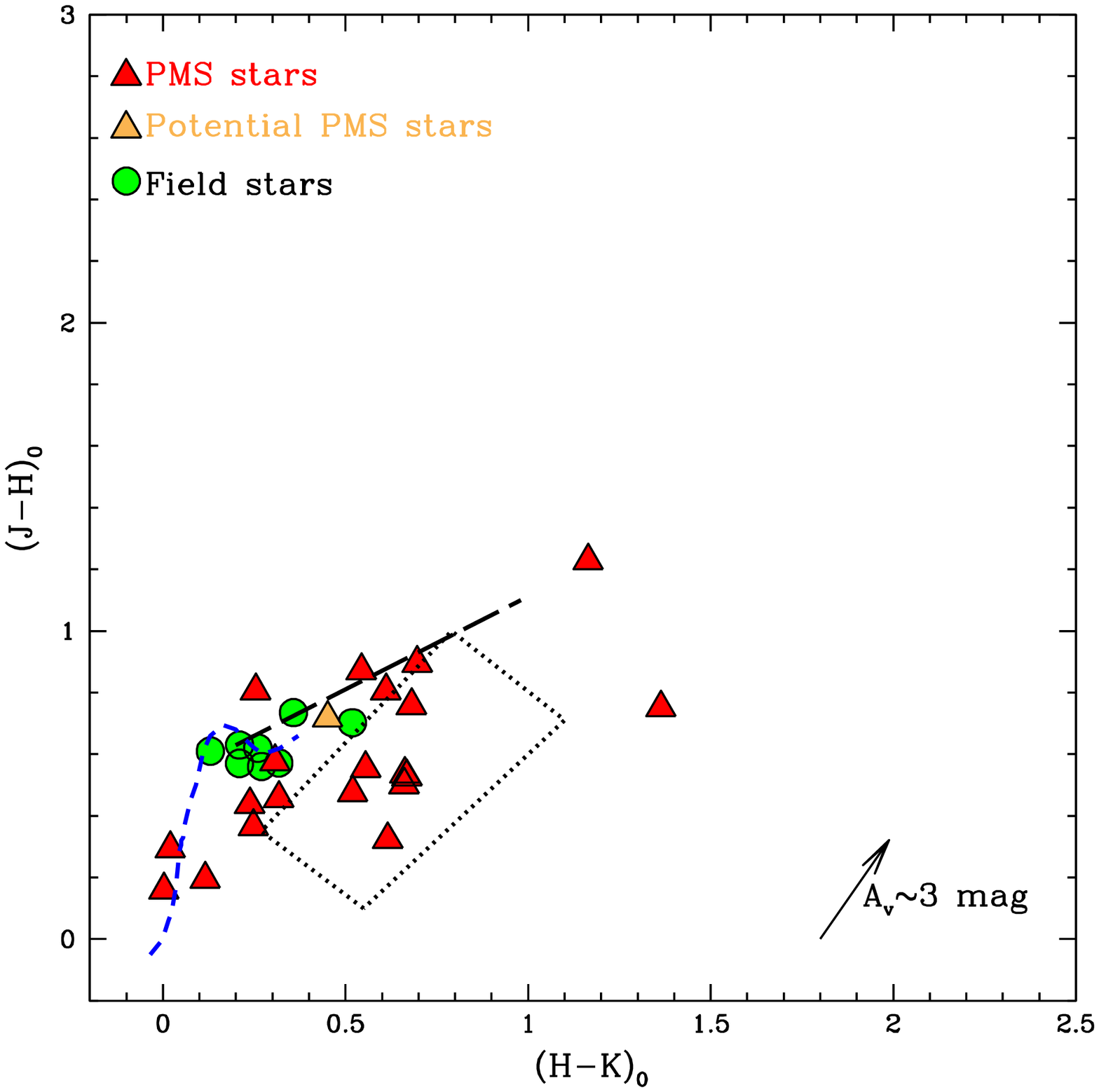}}\\
\end{tabular}
\end{center}
\caption{Near-IR colour-colour diagrams. {\it Left}: observed diagram for the 
identified PMS stars and IRAC sources (grey dots) in Sh~2-284 with $J$, $H$, 
and $K$ measurements. The Class~I and Class~II YSO candidates by P09
are overplotted.  
{\it Right}: extinction corrected diagram of the PMS stars and H$\alpha$ 
emission field stars. Colours  were de-reddened using the extinction values 
reported in the Table~\ref{tab3} and assuming a normal extinction law. The 
rectangle represents the loci of Herbig Ae/Be stars \citep{hernandez05}.
In both diagrams the dashed curve represents the colours of Main Sequence 
stars, while the dashed-dotted line indicates the locus of classical 
T~Tauri stars \citep{meyer97}. The reddening vector is also indicated.}
\label{img:colcol}
\end{figure*}

Near IR colours are useful to select objects with optically thick disks,
typical of Classical T Tauri stars. Fig.~\ref{img:colcol} shows near-IR 
CC diagrams for the PMS stars and for all the IRAC sources that posses
2MASS $J$, $H$, and $K$ measurements. Most of the Class~I and Class~II 
YSO candidates by P09, overplotted on the left panel of 
Fig.~\ref{img:colcol}, have near-IR colours consistent with reddened YSOs, 
but several are scattered towards bluer ($J-H$) colours and fall below the 
locus for classical T~Tauri stars \citep{meyer97}, in a region consistent
with colours of Herbig Ae stars. 
Likewise, many of the PMS stars have colours consistent with those of 
classical T~Tauri stars, but several lie below that locus as well. 

\citet{hernandez05} proposed a criterion to identify candidates to Herbig 
AeBe stars on the {\em intrinsic} near-IR CC diagram. Such diagram for the PMS
stars in Sh~2-284 is shown in the right panel of Fig.~\ref{img:colcol}.
The field H$\alpha$ emission stars are also plotted for comparison.
The figure also shows the approximate locus corresponding to 
intermediate-mass PMS stars \citep[see e.g.][]{hernandez05}. The intrinsic
colours were derived using the spectral types (see Section~\ref{sec_SEDs}).
Seven PMS stars satisfy the criterion by \citet{hernandez05}.

Finally, we notice that a few YSO candidates by P09 have near-IR 
colours similar to those of giants. Recent works \citep[e.g.][]{Oliveira07} 
have pointed out the possible contamination by giants among the samples 
of IRAC-selected YSO candidates.

\subsubsection{Optical colour-magnitude diagrams}
\label{opt_diags}

\begin{figure*}
\begin{center}
\begin{tabular}{cc}
\multicolumn{1}{l}{\includegraphics*[height =8.7cm]{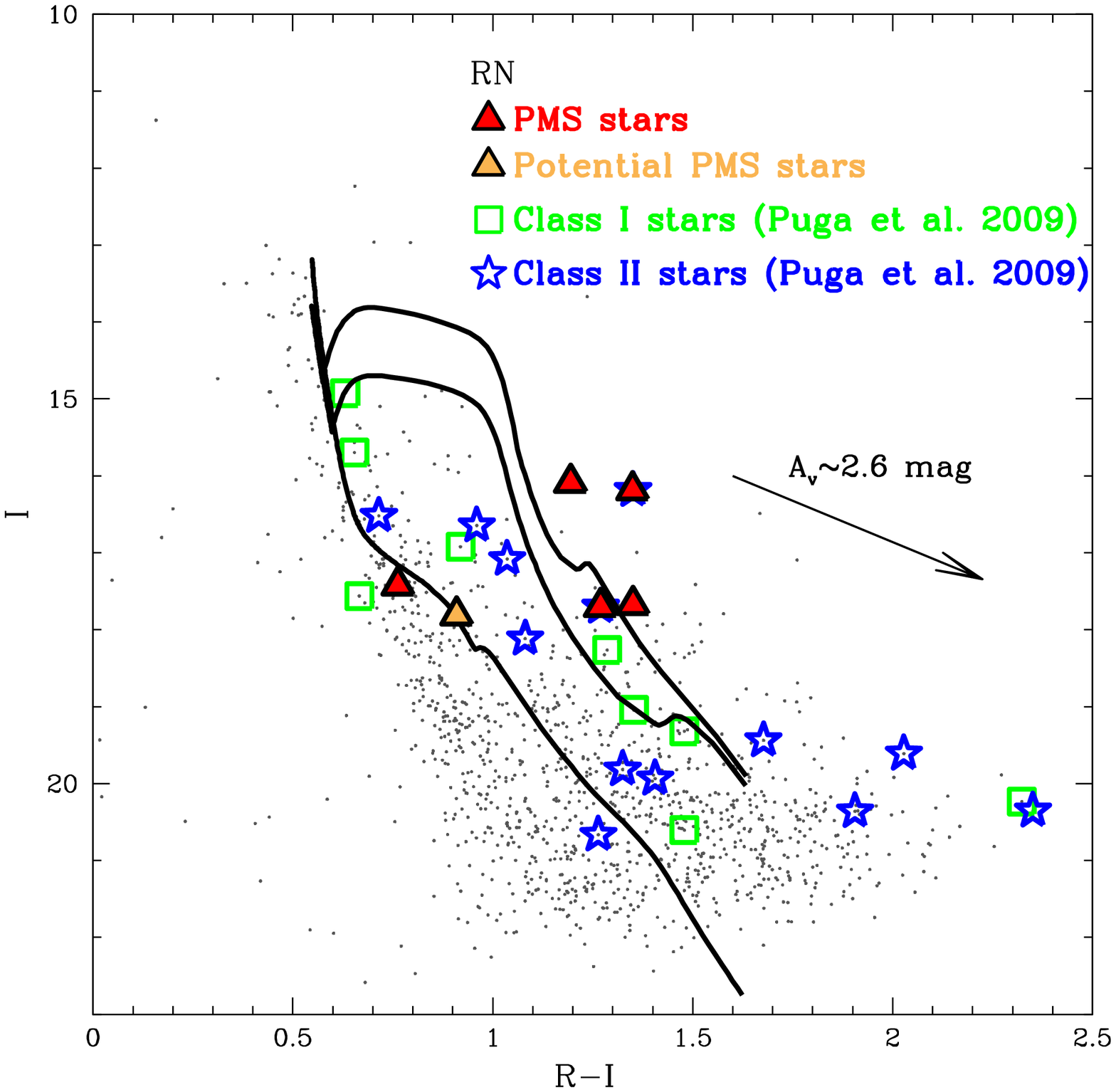}}&\multicolumn{1}{l}{\includegraphics*[height =8.7cm]{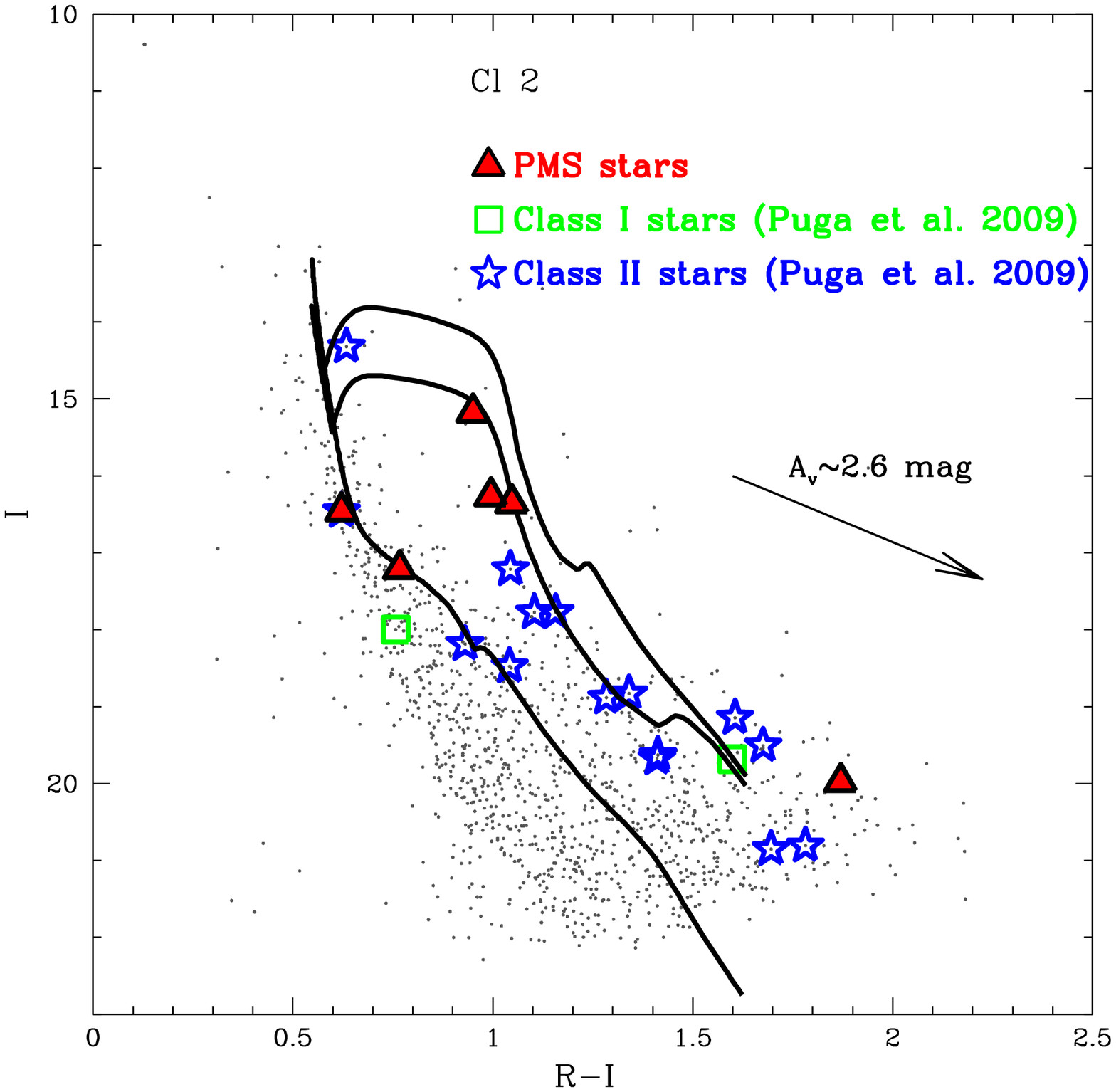}}\\
\multicolumn{1}{l}{\includegraphics*[height =8.7cm]{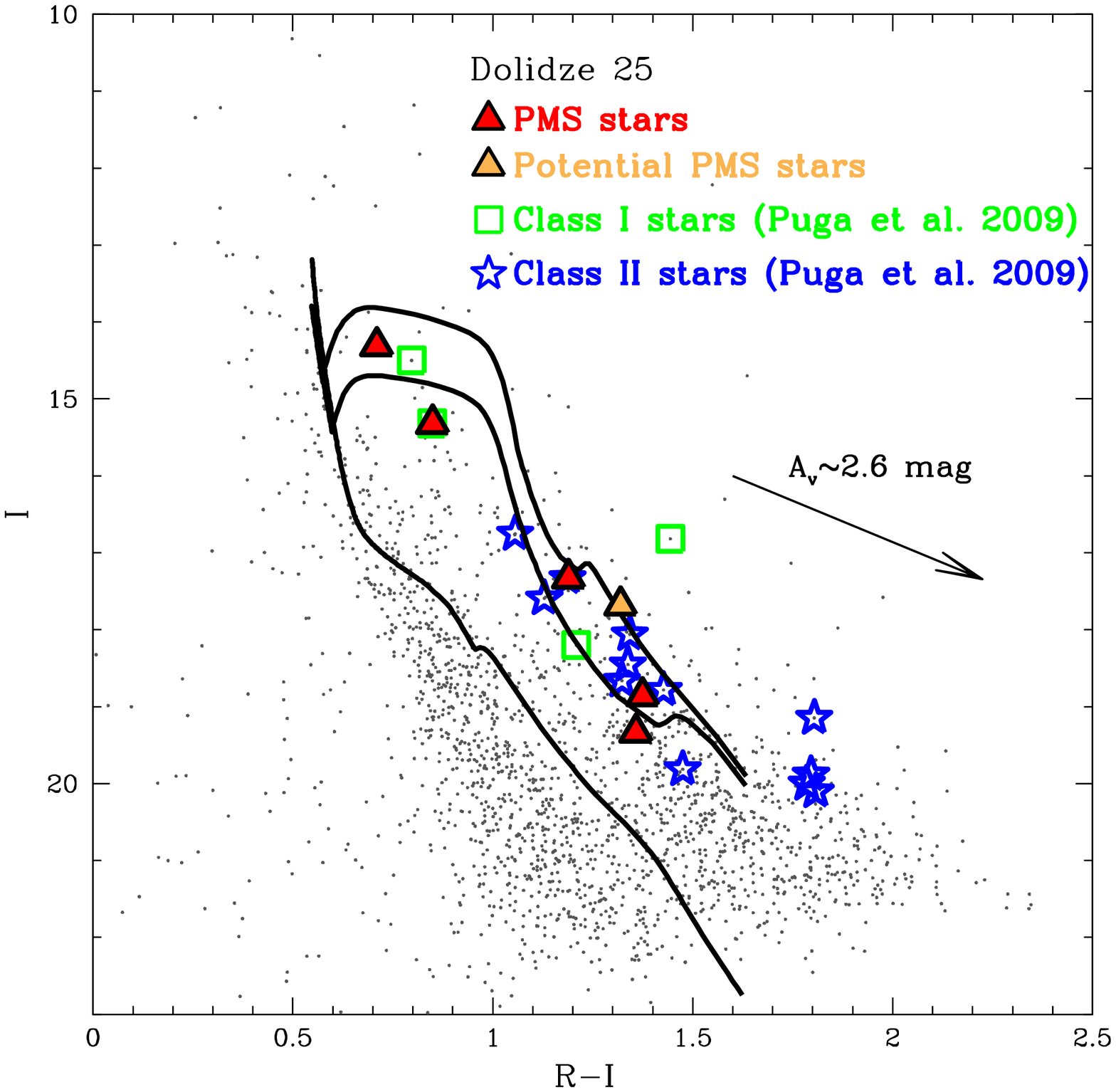}}&\multicolumn{1}{l}{\includegraphics*[height =8.7cm]{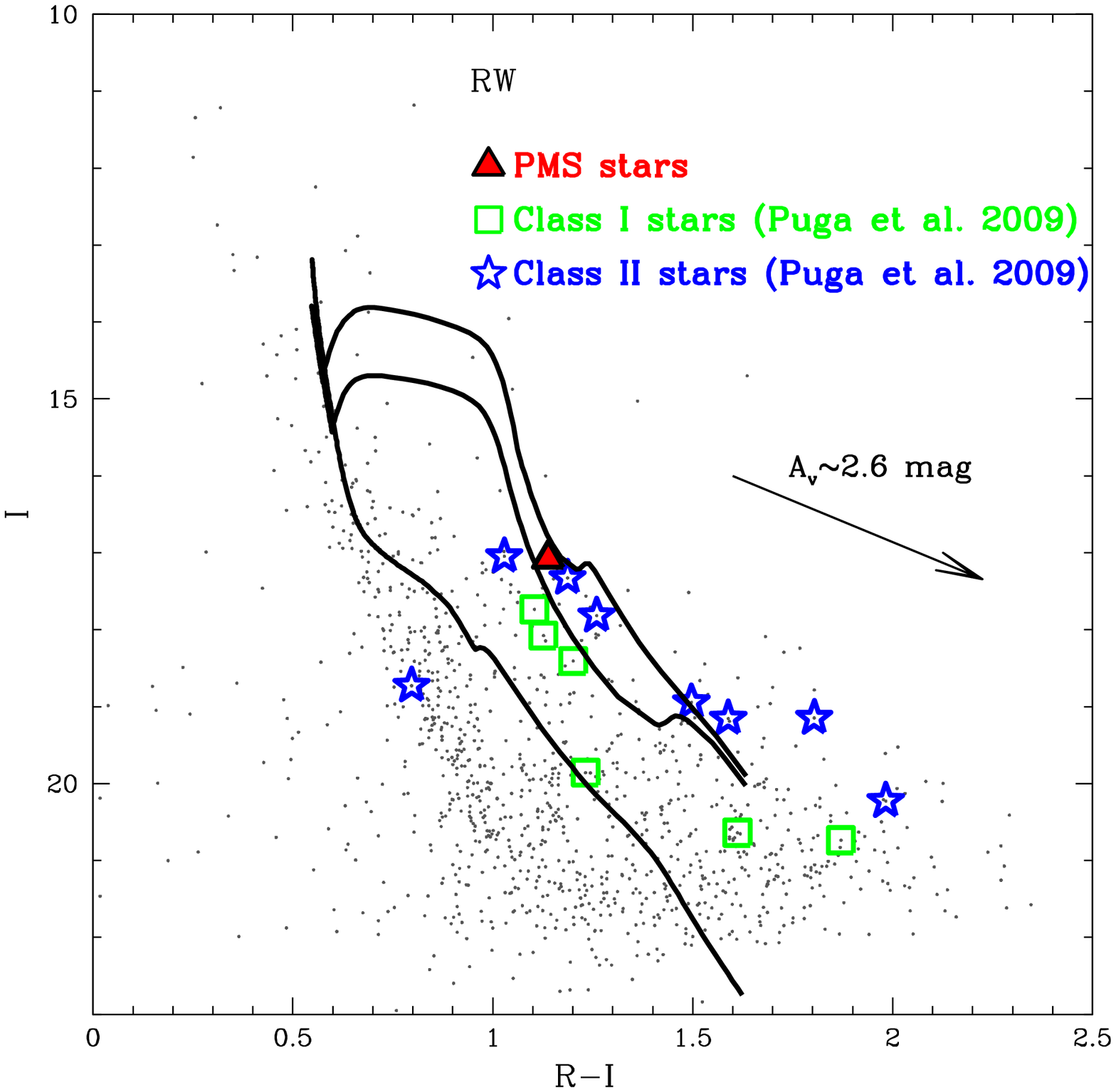}}\\
\end{tabular}
\end{center}
\caption{$I$ vs. $R-I$ diagrams of four regions in Sh~2-284, each covering approximately 
 the same area of $\sim$~40arcmin$^2$ in the sky. The regions correspond to RN, Cl2, 
 Dolidze~25, and RW, as designated by P09.  The 1,2  and 20\,Myr isochrones, 
 with  metallicity Z=0.004 by \citet{tognelli2010}, are represented with solid 
  lines. The \citet{tognelli2010} isochrones were
 computed for  masses in the range 0.2-7 M$_\odot$.  
  Both isochrones were reddened by the average value 
  A$_{\rm v}=$2.6 mag  and shifted by the distance modulus 13.0 mag.}
\label{img:diaRI}
\end{figure*}

The $R$ and $I$ bands are less affected by the continuum dust emission
from the YSO disk. Corrected by interstellar extinction, 
these bands are suitable to prove the continuum stellar photospheric 
contribution. 
Therefore, optical colour-magnitude diagrams can be used to trace the 
stellar sequence of the different aggregates in Sh~2-284. The VIMOS 
observations cover and area of $\sim$ 900 square-arcmin that include four of 
the aggregates studied by P09, namely Dolidze~25, Cl2, RN 
and RW. From our optical catalog we extracted sources on these regions,
with approximately the same projected sky area ($\sim 40$ arcmin$^2$) 
as analysed by P09. Fig.~\ref{img:diaRI} shows the 
observed $I$ vs. $R-I$ diagrams for each one of the four regions. 
The diagrams show also the 1, 2 and 20\,Myr isochrones with  metallicity Z=0.004
\citep[solid line, ][]{tognelli2010}. 
The spectroscopically selected PMS stars and the YSO candidates by 
P09 in these regions are also overplotted with the same 
symbols as in Fig.~\ref{img:diaspit}.

The average extinction, derived from the spectral type of the PMS 
stars (See Section~\ref{sec_SEDs}), is approximately the same in the 
four regions. The 1, 2 and 20\,Myr isochrones shown in Fig.~\ref{img:diaRI} 
were reddened by such average value of  A$_{\rm V}\sim$ 2.6 mag. 
The magnitudes  of the isochrones were also corrected 
by the distance modulus of Sh~2-284  (13.0 mag).

Most of the PMS stars and YSO candidates are consistent with ages of 
the order of 1-2\,Myr, but some are scattered towards younger or older 
ages. In particular, the PMS stars and YSO candidates are more scattered 
in the diagrams of RN and Cl2 regions than in the diagrams of Dolidze~25 
and RW. This may indicate that the objects in Dolidze~25 and RW are 
younger than those in RN and Cl2, in agreement with the findings 
by P09. They concluded that the YSO candidates projected on 
Dolidze~25 are younger, in relative terms, than those in the other 
aggregates. This result is further discussed in Section~\ref{sec_mass}.

\subsubsection{IR colours versus H$\alpha$ emission}
 
\begin{figure}
\begin{center}
\resizebox{\hsize}{!}{\includegraphics[width=4.3cm,height=3.8cm]{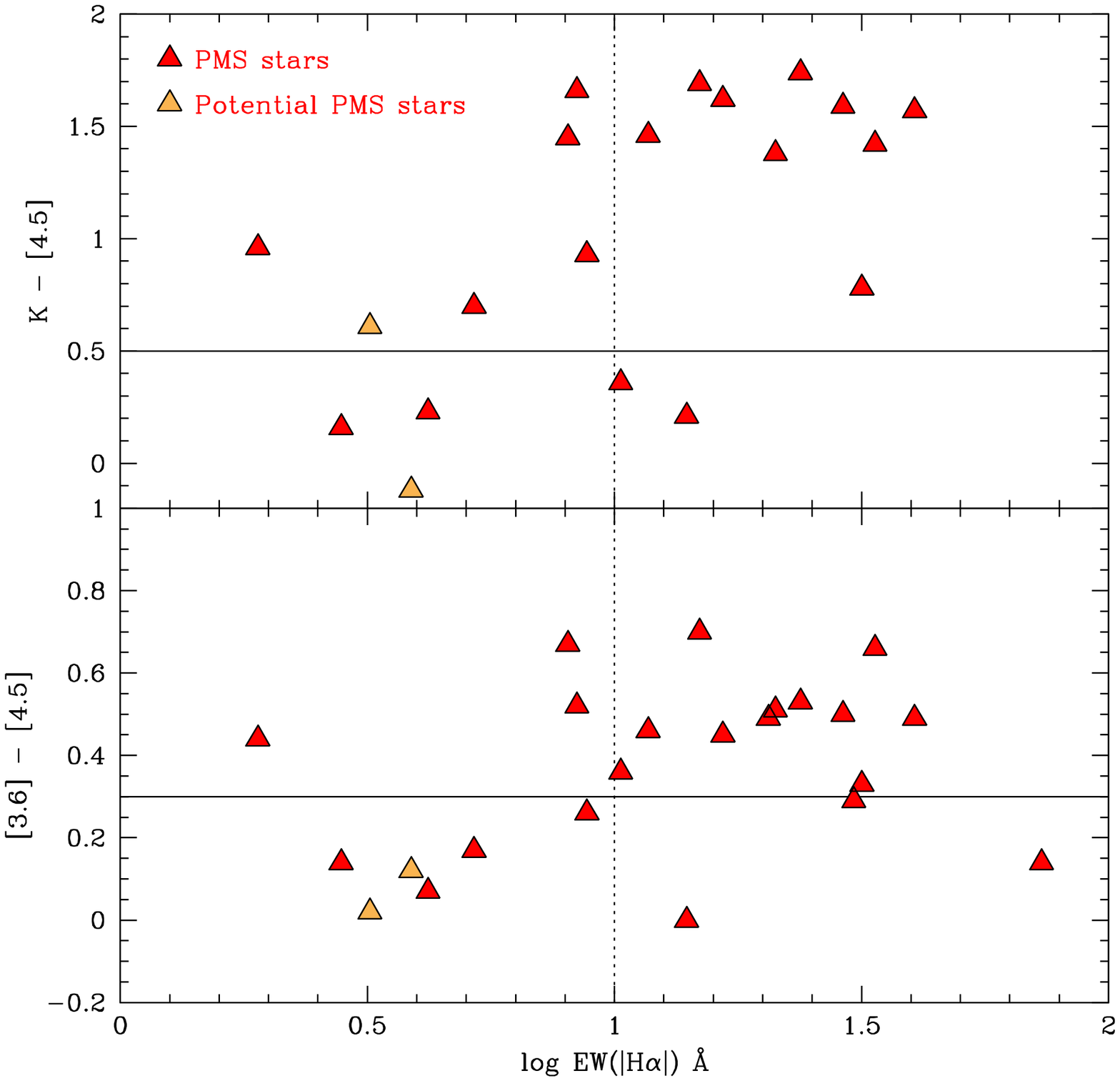}}
\caption{IR colours versus the strength of the H$\alpha$ emission of the
PMS stars in Sh~2-284 . The upper panel shows the $K-[4.5]$ colour v.s. 
EW$_{H\alpha}$ plot, while the lower one the $[3.6]-[4.5]$ colour 
v.s. EW$_{H\alpha}$ plot. The horizontal lines in each panel represent
the minimum limit for IR-excess objects according to \citet{Rob06}, 
i.e. $K-[4.5]>$0.5 and $[3.6]-[4.5]>$0.3 mag, for the upper and 
lower panels respectively.}
\label{WHa_IRCol}
\end{center}
\end{figure}

Previous works \citep{Har98, Lad06} shown that stars with strong 
H$\alpha$ emission in Taurus and IC~348 tend to possess large ($K-L$) 
excess. The relationship between IR colours and H$\alpha$ equivalent 
width, EW$_{H\alpha}$, for the PMS stars in Sh~2-284 is shown in
Figure~\ref{WHa_IRCol}.
Most have IR colours above  the minimum 
for PMS stars with optically thick disks (i.e. $K-[4.5]>$0.5 and 
$[3.6]-[4.5]>$0.3 mag, respectively) according to the models by \citet{Rob06}.
Most of these
 objects, likely having optically thick disks, also possess EW$_{H\alpha}$ above the 
approximate limit for accretion (i.e. 10~\AA). On the other hand, most 
objects with no prominent H$\alpha$ emission have small IR colours.
Thus, consistently with the results of previous works, the IR colour index 
of the PMS stars in Sh~2-284 can be also used as diagnostic for accretion. 
However, the colour indices can be used as overall indicators, but cannot 
replace detailed measurements of accretion for individual objects.

\subsection{Spectral Energy Distributions}
\label{sec_SEDs}

The characterization of the  PMS stars and their circumstellar matter is 
performed also through the analysis of the 
SED \citep{Lad06}. In order to derive the de-reddened SEDs it is necessary 
to use all the available information from the multi-band photometry and 
spectroscopy. The following steps were adopted to determine the SEDs:

\begin{itemize}

\item 
  from the spectral types  the effective temperature $\rm{T_{eff}}$ 
  and the intrinsic colour  (V-I)$_0$ of the stars were derived  by using 
  the conversion tables of \citet{kenyon95}.

\item 
   the reddening E($V-I$) and the absorption in the $V$ band, A$_v$, were 
  derived comparing the observed $V-I$ colour, with (V-I)$_0$, and assuming
  A$_V$/E($V-I$) = 1.92 \citep{cardelli89}. The $R-I$ color was used where the V magnitude was not 
  available.
   Similarly, the absorption in the 
  other bands was calculated. The values of the effective temperature and 
  the visual extinction for each object are reported in Table \ref{tab3}.

\item 
  the magnitudes in all bands were de-reddened and converted into 
  fluxes using the fluxes at zero magnitude given by \citet{bessel79}.

\end{itemize}

The dereddened SEDs were derived for 30 of the 35 objects with spectroscopy 
and H$\alpha$ in emission; the other five objects lack enough photometric 
measurements. Since we do not have U-band photometry, an intrinsic $U$ magnitude 
was derived from the de-reddened $B$ magnitude and the intrinsic $(U-B)_0$ 
colour of the corresponding spectral type.

\begin{figure*}
\begin{center}
\begin{tabular}{cc}
\multicolumn{1}{l}{\includegraphics*[height =9.1cm]{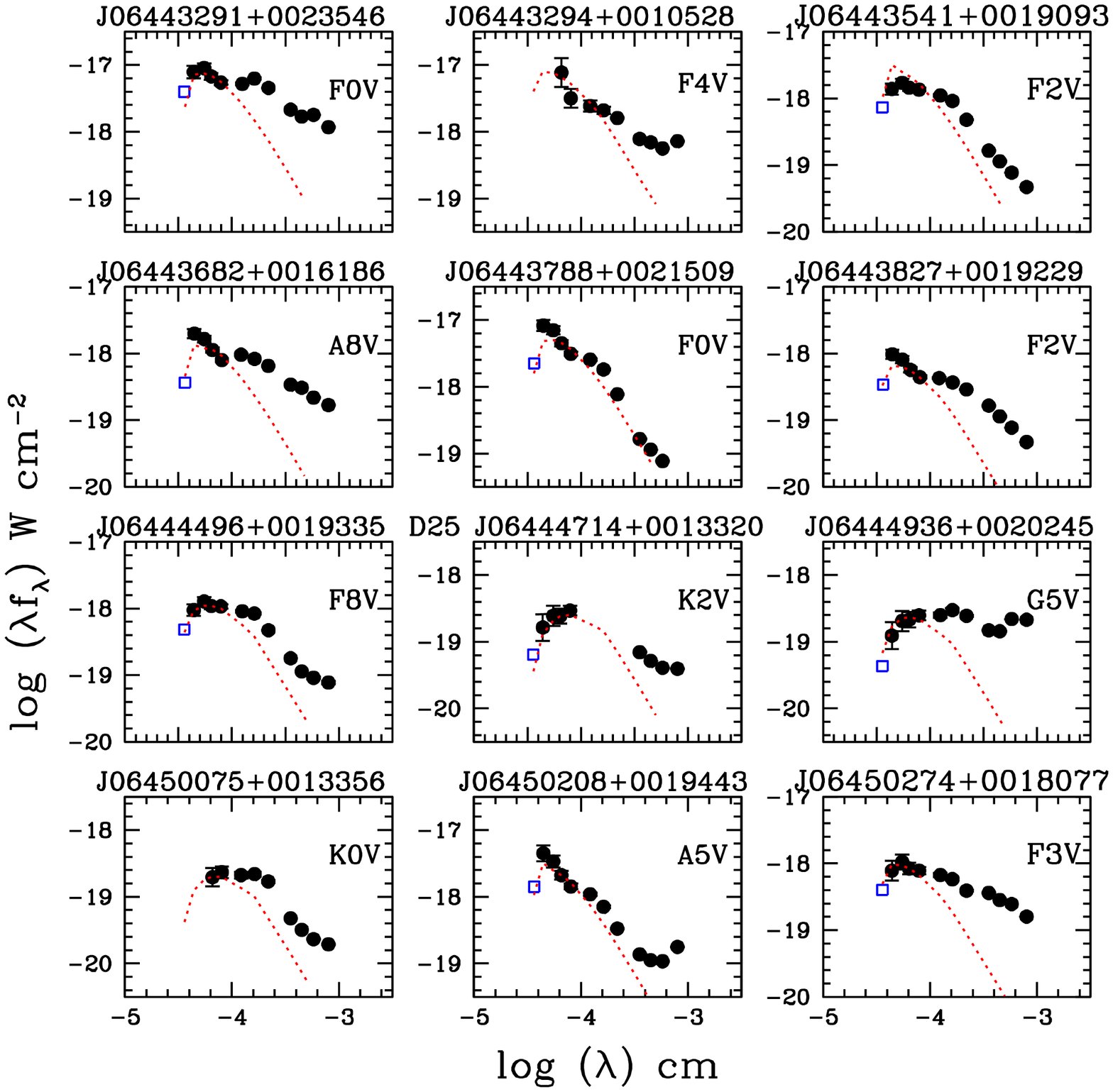}}&\multicolumn{1}{l}{\includegraphics*[height =9.1cm]{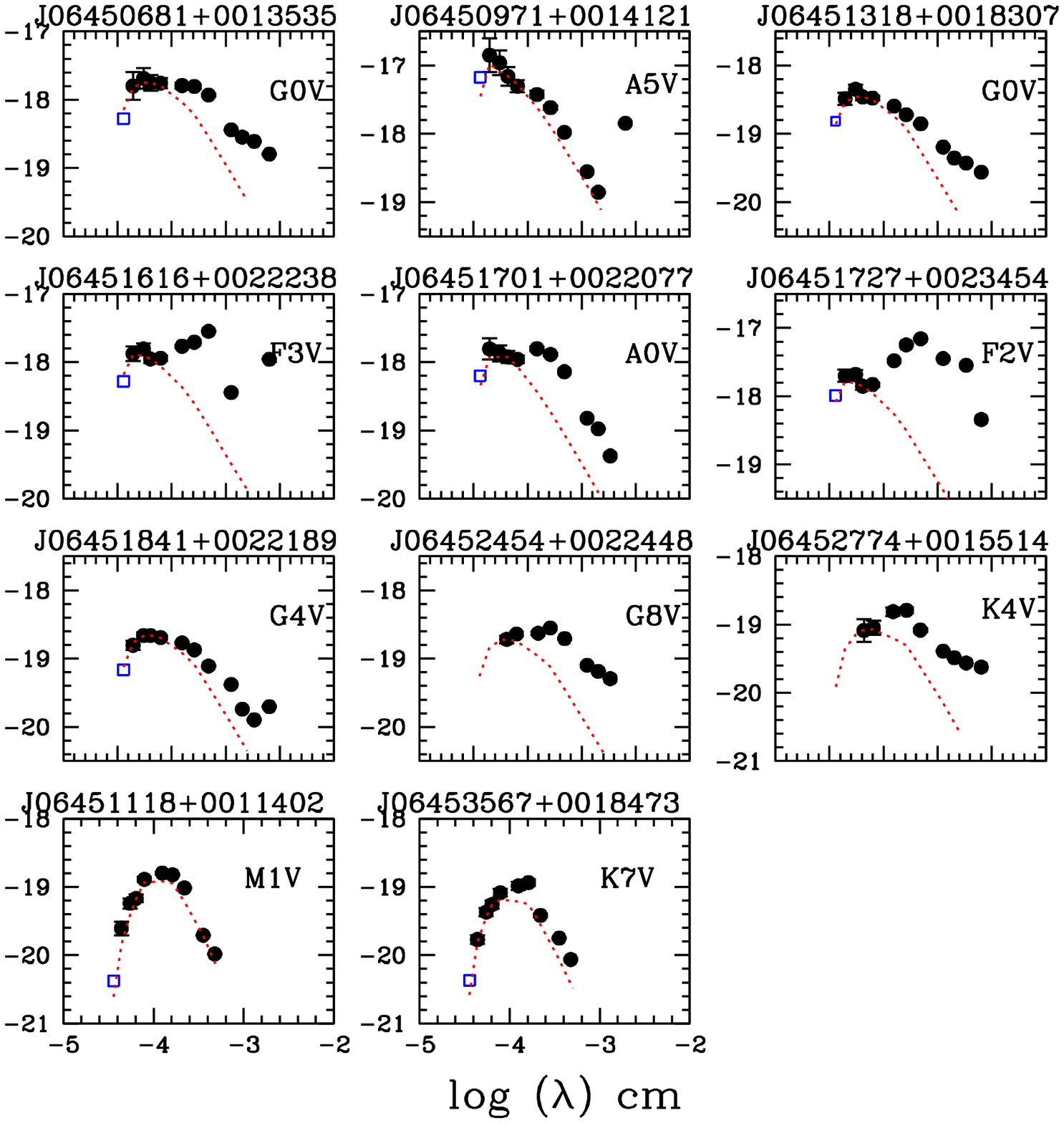}}\\
\end{tabular}
\end{center}
\caption{Spectral energy distribution of PMS stars in Sh~2-284. 
  The dots represent the dereddened flux at each band. The dashed 
  lines represent the SED of normal photospheres of the same spectral 
  type as the object. The blue open squares represent the flux in 
  the U band, calculated from the  B magnitude 
  as explained in the text. On the bottom right the SED for the two potential PMS are plotted.}
\label{img:sed}
\end{figure*}

To further investigate excess emission in the infrared, and thus the 
presence of a circumstellar disk and/or of an envelope, we compared the 
derived SEDs with those of normal photospheres of the same spectral 
type. This is shown in Fig.~\ref{img:sed} for the PMS stars only, where the black dots 
and dotted lines represent the SEDs of the object investigated  and of 
normal photospheres, respectively; the error bars were computed taking 
into account an error of one spectral subclass. Note that the target 
and standard star fluxes were normalized at $R$-band flux. An inspection 
of the figures clearly reveals the stars with IR excess.

\subsection{Luminosities}
\label{sec_lum}

The total luminosity for each object was computed in two ways as follows.
First, by integrating the de-reddened SEDs, using a cubic--spline fit to 
the observed $\lambda$F$_{\lambda}$ ~vs~ $\log\lambda$ distribution assuming 
isotropic radiation and the distance of  4.0~Kpc.  The integration of the SEDs 
was carried out in the interval from $\lambda_B$ to $\lambda_{end}$, the 
terminal wavelength ($\lambda_K$, $\lambda_L$ or $\lambda_M$, depending 
on the available IR photometric data). In order to determine the contribution 
of the flux downward of $\lambda_{end}$, we extrapolated the de-reddened flux 
distribution with a black-body tail, following the method by \citet{cohen1973}.  
The typical errors in $\log\rm{{L_{int}}}$ were estimated to be less than 
about 0.06 dex. Second, the $V$ magnitude corrected for interstellar 
extinction was used to obtain the bolometric luminosity ($\rm{L_{BC}}$), 
by adopting a bolometric correction appropriate to a main--sequence star 
of the same spectral type. The relations given by \citet{kenyon95} were 
used to compute the bolometric corrections.

\begin{figure}
\begin{center}
\resizebox{\hsize}{!}{\includegraphics[width=4.3cm,height=3.8cm]{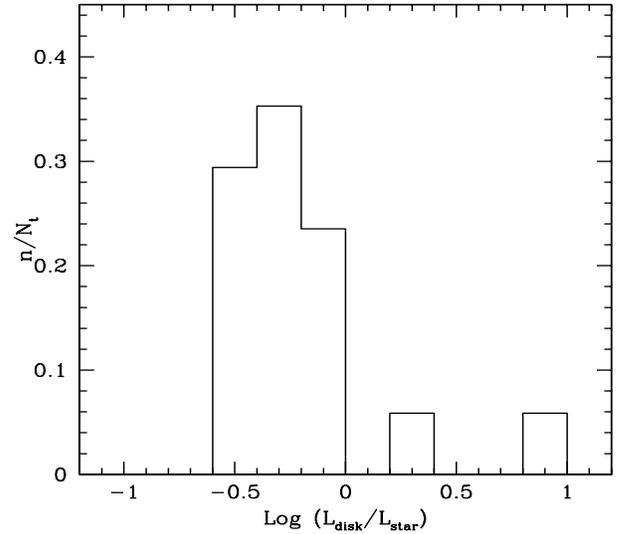}}
\caption{Distribution of the disk-to-photospheric luminosity ratio 
       of PMS stars in Sh~2-284.}
\label{img:isto}
\end{center}
\end{figure}

The difference between the luminosity derived from the integration of 
the SED and that determined from bolometric corrections can be attributed 
to the excess luminosity of the circumstellar disk.
For a flat, optically-thick disk, which merely reprocesses light from the
star, the $\rm{L_{disk}/L_{star}}$ ratio is 0.25 \citep[e.g.][]{All06a}. Such 
value is consistent with the results of recent Spitzer investigations of 
nearby star forming regions \citep[e.g.][]{Har07, Alc08}. 
The distribution of the disk-to-star luminosity for the PMS stars identified
in this work is displayed in Fig.\ref{img:isto}.
The histogram peaks at around $\log\rm{(L_{disk}/L_{star})}=-0.35$, i.e. at 
a disk luminosity which is about half the stellar luminosity and a factor 
of about 2 higher than what found in previous studies of nearby star 
forming regions. Since the cluster is about an order of magnitude more 
distant than the nearby star forming regions, this result reflects the 
fact that our spectroscopic survey revealed the most luminous disks in the 
Sh~2-284 complex. A detailed study of the disk parameters around the PMS 
stars in this region is beyond the scope of the paper, and is deferred to 
a future work. 

The luminosity, $\log\rm{{L_{BC}}}$, derived by adopting a bolometric 
correction should be the best approximation to the photospheric stellar 
luminosity. In the absence of IR excess $\log\rm{{L_{BC}}}$ should be
consistent with $\log\rm{{L_{int}}}$. We have confirmed that for the 
objects without significant IR excess 
$\log\rm{{L_{int}}} - \log\rm{{L_{BC}}}$ $<$ 0.04~dex, which is well 
within the mean error estimated for $\log\rm{{L_{int}}}$. We thus use
the luminosities derived by adopting a bolometric correction for the 
subsequent analysis. Such values are reported in Table~\ref{tab3}.

\subsection{Masses and ages}
\label{sec_mass}

Using the derived effective temperatures and bolometric luminosities it 
is possible to estimate the masses and ages of the PMS stars by comparison 
with PMS evolutionary tracks and isochrones on the HR diagram. 
To this aim we adopted the evolutionary tracks and isochrones by 
\citet{tognelli2010} which are computed for the right metallicity of the 
Sh~2-284 star forming region and that include up-to-date input physics.

\begin{figure}
\begin{center}
\resizebox{\hsize}{!}{\includegraphics[width=5cm]{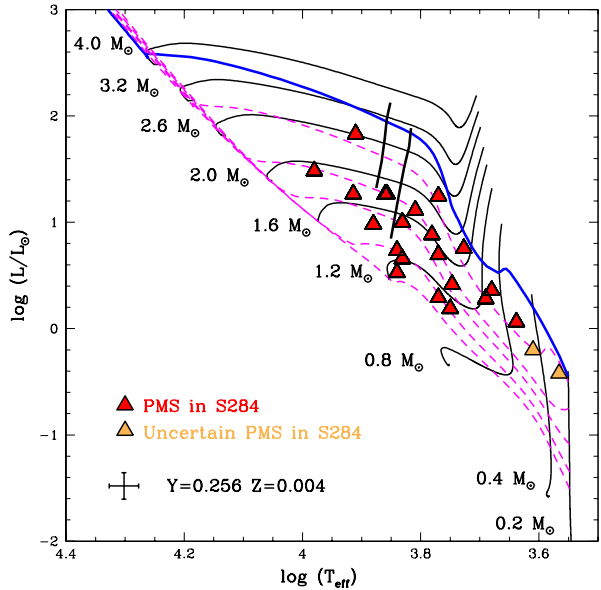}}
\caption{HR diagram of the PMS stars in Sh~2-284. The position of the two 
potential PMS stars  is also
shown in the diagram. The PMS evolutionary tracks, ranging from 0.2~M$_\odot$ 
to 4~M$_\odot$ \citep{tognelli2010}, were computed for the appropriate cluster metallicity Z=0.004.
The 1\,Myr isochrones is represented with the continuous thick line, while 
the 2, 4, 6,  10 and 15\,Myr isochrones are plotted as dashed lines. The instability 
strip is also indicated with the two thick, almost vertical, black lines.}
\label{img:diateo}
\end{center}
\end{figure}

The HR diagram for the PMS stars in Sh~2-284 is shown in Fig.~\ref{img:diateo} 
where the Z=0.004 tracks and isochrones are overplotted. The derived masses 
and ages are reported in Table~\ref{tab3}. Here, the error in mass is half the 
separation between two consecutive evolutionary tracks. The error on the age 
determination is half the separation between two consecutive isochrones. 

The mass of the confirmed PMS stars identified in this work ranges from 
 0.7 to 2.6\,M$_\odot$, while the age from 1.5 to 13\,Myr. 
The lower limit
for the mass is imposed by the limiting magnitude of our spectroscopic
survey, but should the potential PMS candidates be confirmed, this limit 
would drop to about 0.3\,M$_\odot$. The age-spread may be explained in terms 
of different star formation episodes, most likely related to triggered star 
formation in the region (P09). 
Interestingly, most of the PMS stars very well tracing the 2\,Myr 
isochrone (represented by solid lines in Fig.~\ref{img:diateo}), 
lie on the field of Dolidze~25.
Noteworthy, the position on the HR diagram of the two potential PMS stars 
(see Section~\ref{sec_sel_pms}) match quite precisely the  2\,Myr isochrone. 
One of these --the M1-type star-- lies also on the field of Dolidze~25, 
while the other one --the K7-type star-- is located to the north of the 
cluster, in region RN by P09. 
Table \ref{tab3} summarizes  the spatial distribution of the PMS in the Sh 2-284 region.

All this shows that the PMS stars 
in the field of Dolidze~25 have an average age of  $\sim$1-2\,Myr with a very 
small dispersion, and are indeed younger than the PMS stars in the other 
aggregates of Sh~2-284, as previously inferred by the optical CM 
diagrams of Fig.~\ref{img:diaRI} and in agreement with the results by 
P09, based on the relative number of Class~I-to-Class~II 
sources in the different regions.

\subsection{Candidates to PMS $\delta$-Scuti pulsators}

The PMS $\delta$-Scuti candidates were selected based on the positon
of the PMS stars relative to the instability strip. To this aim, we 
overplot the strip by \citet{breger98} on the HR diagram 
(c.f. Fig.~\ref{img:diateo}). Eight stars fall, whitin the errors, in the instability 
strip. These objects are flagged in Table \ref{tab3}.  Four of them 
satisfy the Herbig AeBe criterium by \citet{hernandez05} in the 
near-IR colour-colour diagram. These candidates will be the subject 
of a follow-up study by means of time-series photometry to 
confirm wether or not they are indeed pulsating stars and in that 
case, to derive their pulsational properties (frequencies of
oscillations). These data can be compared with the predictions of
non-radial pulsational models by using asteroseismical techniques 
to estimate independently stellar fundamental parameters such as
luminosity, effective temperature and mass, as well as the distance 
of the host star forming region. 

\section{Discussion}

\subsection{Structure and triggered star formation}
 
 The H{\sc ii} in Sh~2-284, with a radius of about  13.5\,pc, is powered by the 
 high-mass star population in Dolidze~25 (presumably the oldest 
 cluster in the region) and in particular by the O9 type star. P09 
 define the boundaries of this  H{\sc ii} region as ``bubble B1''. 
 The other two clusters detected with IRAC/SPITZER on the boundaries 
 of B1, namely Cl2 to the north-west of Dolidze~25 and and Cl3 to the 
 south-east, power the two compact H{\sc ii} regions B2 and B3 with 
 radii of about  1.5\,pc and 2.7\,pc, respectively. 
 Other rimmed and cometary clouds in the inner 
 regions of B1 have been clearly detected in the IRAC/SPITZER images
 containing a number of YSO candidates; in particular, the ``tail'' of 
 RW apparently points away from Dolidze~25 and contains a number of 
 Class~I sources.
 Such hierarchical structure  and the spatial distribution
 of the different IR-class YSO candidates, indicate that star formation 
 has occurred in the different places of the H{\sc ii} region over a 
 few Myr. 
  
 As  demonstrated in this work, 
 Sh~2-284 is not only a site of high-mass star formation, but its population 
 also comprises very young intermediate and low-mass stars, similarly as 
 in other star forming complexes in which triggered star formation is 
 in act, like in the Orion OB association. 
 
 Spectroscopic surveys in the optical are generally biased 
 towards the detection of Class~II and Class~III sources rather than to 
 Class~I sources. The photometric properties of the Class~I and Class~II
 YSO candidates drawn from the IRAC/Spitzer observations are very similar 
 to those of the spectroscopically confirmed PMS stars.
 Therefore, although there may be some extragalactic contaminants
 specially among the Class~I sources, we can safely conclude 
 that the sample of YSO candidates represents well the population of 
 young stellar objects in Sh~2-284 and can be used to trace the star 
 formation in the region. 
 
 As concluded from the optical colour-magnitude diagrams, different aggregates 
 seem to trace different star formation episodes. In fact, from the relative 
 numbers of  Class~I to Class~II objects, P09 determined that 
 the YSOs projected on the field of Dolidze~25 are younger than those projected
 on  the other aggregates in Cl2 and Cl3. Despite our low-number statistics, 
 the PMS stars projected on the field of Dolidze~25 appear sharply coeval 
 both, on the HR diagram and in the optical colour-magnitude diagram. More 
 importantly, such stars are also significantly younger than the stars in 
 the other aggregates, in line with the  P09 conclusions. 
 Noteworthy, the IRAC YSO candidates on the field of Dolidze~25 also show a 
 very low dispersion along the  2\,Myr isochrone on the $I$ vs. $R-I$ diagram 
 (see Figure \ref{img:diaRI}).
 These facts seem to support indeed the idea of a possible alignment of 
 very young stars in the line of sight, which are located on the near-side 
 of the B1 shell, as suggested by P09 and \citet{delgado2009}.
 In addition, the high relative number of Class~I sources (158) with respect 
 to Class~II sources (188) in Sh~2-284, points towards a non-continuous star 
 formation in the region. 
 We thus conclude that star formation on the B1 shell was most likely
 triggered by the impact of the massive stars of Dolidze~25, leading 
 first to the formation of the stars in clusters Cl2 and Cl3 and then 
 to the stars on the near-side of B1. The size of the two compact H{\sc ii} 
 regions B2 and B3, and the age-spread of the PMS stars in the other 
 aggregates suggests that the Cl2 and Cl3 formed some 2\,Myr before
 the youngest PMS stars appeared on the birth-line. 
     
 \citet{delgado2009} predict two populations of different age in Dolidze~25, 
 one $\sim$4Myr old and the other with an average age of 40\,Myr, but the 
 presence of the  O9-type star in Dolidze~25, implies that the cluster should 
 be younger than $\sim$20\,Myr. With some assumptions on the recombination 
 coefficient, isothermal sound speed and electronic density P09 
 estimated dynamical ages of $<$ 3.8Myr and $<$ 1.8\,Myr for the H{\sc ii} 
 regions B1 and B2, respectively. 
 Considering an age of $\sim$5\,Myr for Dolidze~25 and by comparison with 
 the youngest dynamical estimate, an age-spread of the order of 3 to 4\,Myr 
 would be expected for the region.  However, the low-metallicity of the region 
 may influence the recombination coefficient, isothermal sound speed and 
 electronic density, with an important impact on determinations of the 
 dynamical ages. The age-spread derived from our HR diagram is $<$10\,Myr.
  In addition, although our number statistic is low, we do not find 
  strong evidence of a low-mass star population as old as 40Myr.

 In summary, the first generation of stars in the Sh~2-284 region was 
 formed in a cluster some  10\,Myr ago.
The strong UV radiation and winds of the massive stars swept up the interstellar
matter surrounding the cluster, triggering the formation of the stars associated 
with Dolidze 25. Then, the massive stars of the latter excited the HII region Sh2-284
which in turn prompted the formation of other structures, such as the Cl2 and Cl3 clusters.

Then the youngest generation of stars 
 and rimmed clouds, with their young clusters and aggregates that we 
 see today in the IRAC/Spitzer images, were formed about 2\,Myr ago. 
 Many Class~I and Class~II YSOs in the region should belong to this 
 latter generation, but many older YSOs of previous generations may 
 have preserve their disks/envelopes.
  
\subsection{Survival of disks/envelopes}
 
 Metallicity may play an important role on circumstellar 
 disk/envelope evolution. In a recent study of the young (age$\sim$1\,Myr) 
 stellar populations in low-metallicity regions of the galaxy, \citet{Yasui09} 
 find that the disk fraction is significantly lower than in star forming 
 regions of solar-metallicity. They suggest that the low-metallicity may 
 induce an increase of the mass accretion rate and/or a very efficient 
 photo-evaporation of the circumstellar matter, both processes leading 
 to a very rapid disk dispersal. As a consequence most stars forming in 
 a low-metallicity environment should experience the disk dispersal at 
 an earlier stage ($<$1\,Myr) than those forming in a solar metallicity 
 environment ($\sim$5-6\,Myr). 
 Other studies of candidates to low-mass PMS stars in the LMC \citep{Spezzi10} 
 also suggest a higher mass accretion-rate, by up to an order of magnitude, 
 in comparison with the values derived for their Galactic counterparts.  
  
 Because of the low-metallicity environment in Sh~2-284, a low fraction of 
 disks/envelope systems among the oldest YSOs is expected. Yet, the large 
 number of Class~I and Class~II YSO candidates detected in Sh~2-284 and the 
 age spread of more than 4\,Myr indicate that a significant fraction of 
 YSOs may have preserved their circumstellar disks/envelope, unless
 they represent the youngest generation. Indeed, the YSO candidates were 
 selected based on their red IRAC colours and the sample might be biased 
 towards the youngest IR-excess objects.
 
 Our VIMOS spectroscopic survey was designed to include not only IR-excess
 YSO candidates, but also PMS candidates that may or may not have IR excess.
 The VIMOS sample is thus unbiased towards IR excess and hence, can be used 
 to investigate how many of the oldest PMS stars have eventually preserved 
 their disks. Of the 23 PMS stars, there are 16 (i.e. $\sim$ 70\%) with an age 
 $t \geq$ 3\,Myr. Of these,  10 posses strong IR-excess, indicative of an 
 optically thick circumstellar accretion disk (see Fig.~\ref{img:sed}). 
 Likewise, from Figure~\ref{WHa_IRCol} we find that  7 out of 13  objects with
 $t \geq$ 3\,Myr show strong IR-excess and hence, should possess an optically 
 thick disk according to the IR-colour criterium by \citet{Rob06}.
 These results would also indicate that a significant fraction of 
 the oldest PMS stars in Sh~2-284 have preserved their accretion disks, 
 despite the low-metallicity environment. The number statistics of our 
 sample is low, but should this result be confirmed with further 
 spectroscopic observations, it will imply that disk survival may not only
 depend on metallicity, but also on other environmental physical conditions 
 and/or the properties of the central objects.
  
 Assuming that all Class~II sources eventually evolve into Class~III sources 
 when their disks are fully accreted by the PMS stars, \citet{Bertout07} 
 estimated that the disk lifetimes for the Taurus-Auriga PMS population 
 should be given by $t_{disk}=4 \times (M_{\star}/M_{\odot})^{0.75}$\,Myr.
 This means that the disk of a 0.5\,M$_{\odot}$ Taurus-Auriga star
 survives $\sim$2\,Myr, on the average. Though applicable to solar
 metallicity stars, such relationship would imply an average disk-lifetime 
 of the order of  5.5\,Myr for the average  $\sim$1.5\,M$_{\odot}$ star in our 
 PMS sample, i.e. rather consistent with the average age of 5.3 Myr drawn from the HR diagram. 
 However, our sample is not complete specially in the low-mass regime 
 and hence, may be biased towards massive PMS stars.
  In addition, the relationship by  \cite{Bertout07} was derived
 for young stars with masses sensibly lower than those in Sh-2-284.
 Hence, these results should be taken with care. In any case, a complete
 spectroscopic survey of the region, down to about 0.3 M$_\odot$, is crucial
 to investigate the disk survival in the region.
This calls for a more complete spectroscopic survey. 

\section{Conclusions}

Through VIMOS$@$VLT, 2MASS and Sptizer data  we have investigated
the low-metallicity star forming region Sh~2-284.
On the basis of H$\alpha$ emission, IR excess, as well as of IR colour-magnitude
and colour-colour diagrams, we identified 23 new PMS stars in the region. 
We have characterized the star sample by determining the PMS stellar 
physical parameters. The IR properties of these stars and the Spitzer/IRAC 
selected YSO candidates by P09 are similar in many respects,
but the latter sample is biased towards IR-excess objects.

 By detailed spectroscopic parallax of three OB stars in the region we have estimated a distance of 4~Kpc for 
Sh 2-284.
  
The PMS stars found in the field of view of Dolidze~25 have an average age 
of  2\,Myr, i.e. younger than the objects in the other aggregates of the region,
in agreement with the results from the number ratio of Class~I--to--Class~II
sources by P09 throughout the region. The age spread of the PMS, 
their spatial distribution and the cloud structure of Sh~2-284 suggest a 
sequence of star formation events, hence, that triggered star formation is 
in act in the region, with the massive stars of Dolidze~25 being the main 
triggers of the B1 shell.
All this supports the idea of a possible alignment of the very young stars 
in the line of sight, which are located on the near-side of the B1 shell, 
as suggested by P09 and \citet{delgado2009}.

Despite the low-metallicity environment, a significant fraction of 
the oldest PMS stars in Sh~2-284 have preserved their accretion disks/envelopes. 
Although this result may be affected by the poor statistics of our spectroscopic 
sample, the large number of Class I and Class II sources in the different 
aggregates and their apparent age spread in the colour-magnitude diagrams 
seems to support it.

Among the 23 discovered young stars in Sh~2-284, we selected  8
with intermediate mass that are very good candidates for PMS $\delta$~Scuti 
type pulsation. These stars will be studied with asteroseismological 
techniques from the ground.
\begin{table*}
\scriptsize  
\caption{Photometry of PMS stars in Sh~2-284.
    In the first column the 2MASS identification is given.
    D25  id indicates that the object is not in 2MASS;
    the second and third columns list the right ascension
    and declination as given by the 2MASS catalogue; in the fourth and fifth columns the
    $B$ and $V$ magnitudes, as extracted from the EXOdat catalogue,
    are reported; the sixth and seventh columns list the
    $R$ and $I$ magnitudes as measured from our VIMOS
    observations; in the eighth, ninth and tenth columns,
    the $J$, $H$ and $K$ magnitudes, as extracted
    from the 2MASS catalogue, are reported; the last four
     columns report the magnitudes in the four IRAC bands (P09).}   
\begin{tabular}{c c c c c c c c c c c c c c}
\hline\hline 
id$^1$ & $\alpha$ (J2000)& $\delta$ (J2000)  & $B$ & $V$ & $R$ & $I$ & $J$ & $H$ & $K$ & $[3.6]$ & $[4.5]$ & $[5.8]$ & $[8]$ \\ 
\hline 
 & & & & & & & & & & & & & \\
J06443291+0023546     & 06 44 32.93  & +00 23 54.9  & 19.53 & 17.79  & 17.02  &  15.76 & 13.39 &  12.11  & 11.22 &  10.01  &   9.48   &   8.67  &  7.82\\
 J06443294+0010528    & 06 44 32.94  & +00 10 52.8  &    -  & -      & 18.21  &  17.07 & 14.01 &  13.12  & 12.23 &  11.21  &  10.44   &   9.85  &  8.89\\
 J06443541+0019093    & 06 44 35.40  & +00 19 09.5  & 19.87 & 18.37  & 17.40  &  16.35 & 14.74 &  13.96  & 13.52 &  13.36  &  13.29   &  12.91  & 12.72\\
J06443682+0016186     & 06 44 36.82  & +00 16 18.3  & 18.22 & 17.30  & 16.98  &  16.32 & 14.58 &  13.83  & 13.03 &  11.89  &  11.37   &  10.94  & 10.19 \\
 J06443788+0021509    & 06 44 37.89  & +00 21 50.9  & 18.33 & 16.96  & 16.30  &  15.35 & 13.87 &  13.22  & 12.99 &  12.97  &  12.83   &  12.67  &  - \\
J06443827+0019229     & 06 44 38.26  & +00 19 23.1  & 18.27 & 17.53  & 17.08  &  16.46 & 15.31 &  14.61  & 13.85 &  12.85  &  12.39   &  12.06  & 11.60  \\
J06444496+0019335     & 06 44 44.95  & +00 19 33.7  & 19.70 & 18.23  & 17.26  &  16.26 & 14.83 &  13.96  & 13.48 &  12.81  &  12.55   &  12.04  & 11.23 \\
 D25 J06444602+0019182 & 06 44 46.02  & +00 19 18.2  &  -    &  -     & 21.85  &  19.98 &   -   &      -  & -     &  15.71  &  15.57   &  -      &   -  \\  
D25 J06444714+0013320 & 06 44 47.14  & +00 13 32.0  & 20.79 & 19.38  & 18.51  &  17.33 & -     &  -      & -     &  13.64  &  13.15   &  12.72  & 11.82\\
J06444936+0020245     & 06 44 49.39  & +00 20 24.8  & 19.90 & 18.63  & 17.97  &  17.21 & 15.79 &  14.77  & 14.00 &  13.01  &  12.31   &  11.08  & 10.13 \\
 J06450075+0013356    & 06 45 00.75  & +00 13 35.6  &   -   & -      & 20.24  &  18.86 & 16.27 &  15.31  & 14.51 &  14.06  &  13.73   &  13.46  & 12.67 \\ 
 J06450208+0019443    & 06 45 02.09  & +00 19 44.3  & 18.22 & 17.18  & 16.56  &  15.77 & 14.62 &  14.13  & 13.84 &  13.06  &  12.39   &  11.46  &  9.81  \\
J06450274+0018077     & 06 45 02.75  & +00 18 07.7  & 20.86 & 19.18  & 18.66  &  17.54 & 15.36 &  14.51  & 13.77 &  13.01  &  12.35   &  11.69  & 10.67 \\
J06450681+0013535     & 06 45 06.81  & +00 13 53.5  & 18.01 & 16.82  & 16.17  &  15.32 & 13.96 &  13.11  & 12.38 &  11.86  &  11.42   &  10.79  &  9.58 \\
J06450971+0014121     & 06 45 09.85  & +00 14 12.5  & 16.35 & 15.44  & 15.05  &  14.34 & 13.15 &  12.71  & 12.54 &  12.33  &  12.33   & -       &  8.07  \\
 J06451318+0018307    & 06 45 13.12  & +00 18 29.6  & 21.48 & 19.85  & 19.21  &  18.08 & 16.34 &  15.67  & 14.85 &  13.77  &  13.28   &  12.85  & 12.25\\
 J06451616+0022238    & 06 45 16.17  & +00 22 24.2  & 21.41 & 19.65  & 18.99  &  17.67 & 14.59 &  13.37  & 11.73 &  12.05  &  -       &  -      &  8.34\\
J06451701+0022077     & 06 45 16.99  & +00 22 07.6  & 19.35 & 18.17  & 17.28  &  16.09 & 14.23 &  13.47  & 12.99 &  12.99  &  12.63   &  12.87  &  -\\
 J06451727+0023454    & 06 45 17.26  & +00 23 45.3  & 19.34 & 18.05  & 17.54  &  16.19 & 13.52 &  11.96  & 10.60 &   9.43  &   8.98   &   8.02  &  7.32 \\
 J06451841+0022189    & 06 45 18.40  & +00 22 19.0  & 19.63 & 18.55  & 18.15  &  17.42 & 16.21 &  15.63  & 15.24 &  14.71  &  14.54   &  14.18  & 12.70 \\ 
J06452454+0022448     & 06 45 24.54  & +00 22 44.8  &  -    &   -    & 18.97  &  17.70 & 16.17 &  15.05  & 14.36 &  13.49  &  12.98   &  12.59  & 11.86 \\
D25 J06452476+0013360 & 06  45 24.76 & +00 13 36.0  &    -  &  -     & 20.69  &  19.33 & -     &  -      &  -    &  15.75  &  15.46   &   -     &  -  \\
J06452774+0015514     & 06 45 27.74  & +00 15 51.5  &   -   &   -    & 20.22  &  18.93 & 16.75 &  15.73  & 15.35 &  14.26  &  13.76   &  13.83	& 12.33	 \\   
  J06451118+0011402  & 06 45 11.19  & +00 11 40.7  & 21.22  & 19.67 & 19.00  & 17.68  & 16.19  & 15.44  & 14.97 &  15.21  &  15.09   &  -       & -          \\
J06453567+0018473    & 06 45 35.70  & +00 18 47.1  & 20.85  & 19.38 & 18.72  & 17.81  & 16.49  & 15.61  & 15.90 &  15.31  &  15.29   &   -      & -          \\
   & & & & & & & & & & & & &\\  
\hline 
$^1$ 2MASS 
\label{tab1} 
\end{tabular}
\vspace{1cm}
\caption{Photometry of field stars in  Sh~2-284 that show emission in the H$\alpha$ line.}
\begin{tabular}{c c c c c c c c c c c c c c}
\hline\hline 
id$^1$ & $\alpha$ (J2000)& $\delta$ (J2000)  & $B$ & $V$ & $R$ & $I$ & $J$ & $H$ & $K$ & $[3.6]$ & $[4.5]$ & $[5.8]$ & $[8]$ \\ 
\hline 
 & & & & & & & & & & & & & \\
J06445227+0014300    & 06 44 52.23  & +00 14 30.0  & 18.92  & 17.39 & 16.70  & 14.97  & 13.24  & 12.68  & 12.41 &  12.20  &  12.18   &   12.05  &   12.00    \\
 J06445433+0021190   & 06 44 54.29  & +00 21 19.0  & 20.98  & 19.24 & 18.05  & 15.94  & 14.09  & 13.48  & 13.14 &  12.90  &  12.88   &   12.67  &  -         \\
 J06450151+0008493   & 06 45 01.51  & +00 08 49.3  &   -    &   -   & 18.98  & 17.49  & 16.18  & 15.57  & 15.44 &  15.21  &  15.18   &   -      &  -         \\
D25 06452246+0015408 & 06 45 22.46  & +00 15 40.8  &   -    & -     & 20.44  & 18.64  &  -     &   -    &  -    &  15.81  &  15.76   & -        & -          \\
J06453429+0022261    & 06 45 34.30  & +00 22 26.5  & 17.47  & 16.07 & 15.41  & 14.27  & 13.09  & 12.46  & 12.25 &  12.29  &  12.29   &  11.97   &  11.95     \\
J06453550+0024158    & 06 45 35.52  & +00 24 16.5  &  -     &   -   & 20.05  & 18.21  & 16.46  & 15.62  & 15.02 &  15.23  &   -      & -        & -          \\
J06453643+0011254    & 06 45 36.45  & +00 11 25.6  & 20.75  & 19.31 & 18.52  & 16.84  & 15.02  & 14.45  & 14.24 &  14.06  &  14.04   &  13.78   &  -         \\
D25 J06454297+0011461& 06 45 42.97  & +00 11 46.1  & 21.11  & 19.75 & 19.22  & 18.24  &  -     &   -    & -     &  15.99  &  15.95   &  -       & -          \\ 
J06454948+0011347    & 06 45 49.45  & +00 11 35.1  & 18.16  & 16.63 & 15.94  & 14.34  & 12.74  & 12.12  & 11.86 &  11.71  &  11.75   &   11.67  &  11.56     \\
J06455486+0008134    & 06 45 54.86  & +00 08 13.4  &   -    & -     & 19.90  & 18.27  & 16.59  & 15.80  & 15.41 &  15.44  &  15.39   &  -       &   -        \\
 & & & & & & & & & & & & &\\  
\hline  
\end{tabular}
\label{tab2} 
\end{table*}
\begin{table*}
\scriptsize
\caption{Summary of spectroscopic information and physical parameters 
    for PMS stars. The errors on
    the visual extinction are in average of a tenth percent using the extinction law given in \citet{cardelli89}. 
    The error on the temperature comes from the one spectral sub-class
    uncertainty on spectral classification. 
    The error in luminosity is dominated by the uncertainty in distance,
    which is of the order of 10~$\%$. For each star we  assumed a distance of  4.0 Kpc. The error in mass  and age is half 
    the separation between two consecutive  tracks and isochrones, respectively. In the last column some notes are given.}
\label{tab3} 
\begin{tabular}{c c c c c  c c c c c l}
\hline\hline 
id   & SpT & EW($\rm{H_\alpha}$)   & EW(Ca I+ Fe I ) &  EW(Na I) & Av   & $\log \rm{T_\mathrm{eff}}$  & $\log \rm{L_{BC}/L_\odot}$ & Mass &age   &  Notes \\
     &     &   [\AA]               &  [\AA]          & [\AA]     &(mag) &                             &                            & (M$_\odot$)  & (Myr)    & \\
\hline 
\hline 
 & & & & & & & & & &\\
J06443291+0023546   &    F0V   &  $-23.80\pm0.40$   &   1.40$\pm$0.10 & 3.30$\pm$0.20      & 3.10  &   3.860   &   1.27   & 1.8  &  4.0  & 1, 3  \\
J06443294+0010528      &    F4V   &  $-139.28\pm1.23$  &      noise	  & 1.70$\pm$0.23  & 3.64  &   3.809   &   1.11   & 1.6  &   4.0  & 1, 3, 7	 \\ 
J06443541+0019093     &    F2V   &  $-4.20\pm0.20$    &   1.45$\pm$0.05 & 2.70$\pm$0.15    & 3.00  &   3.831   &  1.00    & 1.6  &  6.0  & 1, 5    \\
J06443682+0016186     &    A8V   &  $-8.40\pm0.20$    &   0.00$\pm$0.12 & 2.32$\pm$0.20    & 1.90  &   3.880   &  0.98    & 1.4  &  8.0  & 1, 2    \\ 
J06443788+0021509       &    F0V   &  $-2.80\pm0.10$    &   0.70$\pm$0.10 & 2.25$\pm$0.10  & 2.28  &   3.857   &   1.27   & 1.8  &  4.0  & 1, 5      \\
J06443827+0019229       &    F2V   &  $-11.70\pm0.25$   &   0.06$\pm$0.02 & 1.60$\pm$0.10  & 1.30  &   3.831   &  0.66    & 1.2  &  10.0  & 3, 5  \\
J06444496+0019335       &    F8V   &  $-8.80\pm0.15$    &   1.48$\pm$0.05 & 2.12$\pm$0.20  & 2.52  &   3.781   &  0.88    & 1.6  &  4.0  & 5		  \\
D25 J06444602+0019182    &    G0V   &  $-73.20\pm0.90$   &      noise	  &   noise 	   & 5.63  &   3.770   &  0.70    & 1.6  &  5.0  & 5		  \\
D25 J06444714+0013320   &    K2V   &  $-20.50\pm0.50$   &   3.70$\pm$0.20 & 3.60$\pm$0.20  & 2.15  &   3.680   &  0.36    & 1.0  &  1.5  & 3, 4  	  \\
J06444936+0020245       &    G5V   &  $-14.87\pm0.30$   &   2.62$\pm$0.10 & 3.50$\pm$0.20  & 1.70  &   3.747   &  0.42    & 1.2  &  5.0  & 5		   \\
J06450075+0013356       &    K0V   &  $-31.69\pm0.50$   & 	-	  & 3.54$\pm$0.30  & 1.89  &   3.690   &  0.28    & 1.2  &  2.0  & 4		   \\
 J06450208+0019443      &    A5V   &  $-8.05\pm0.10$    &   0.56$\pm$0.05 & 1.95$\pm$0.15  & 2.44  &   3.914   &   1.27   & 1.8  &  5.0  & 2		  \\ 
J06450274+0018077       &    F3V   &  $-33.70\pm0.25$   &   1.30$\pm$0.15 & 2.50$\pm$0.23  & 3.14  &   3.840   &  0.74    & 1.2  &  10.0  & 1, 3        \\
J06450681+0013535       &    G0V   &  $-1.90\pm0.10$    &   0.00$\pm$0.20 & 2.00$\pm$0.10  & 2.00  &   3.770   &   1.25   & 2.3  &  1.5  & 2, 4 	  \\
J06450971+0014121*      &    A5V   &  $-14.00\pm0.20$   &     -           & $2.30\pm0.30$  & 2.11  &   3.910   &   1.83   & 2.6  &  2.0  & 4		 \\
J06451318+0018307       &    G0V   &  $-40.50\pm0.50$   &     -	          & 3.40$\pm$0.20  & 2.64  &   3.770   &  0.29    & 1.0 &   10.0  & 3		 \\
J06451616+0022238       &    F3V   &  $-5.30\pm0.25$    &   0.35$\pm$0.10 & 3.17$\pm$0.30  & 3.01  &   3.840   &  0.53    & 1.2  &  12.5  & 1, 6     \\
 J06451701+0022077      &    A0V   &  $-10.30\pm0.40$   &   0.00$\pm$0.10 & 1.56$\pm$0.20  & 3.82  &   3.980   &   1.48   & 2.0  &  4.0  & 6		 \\
 J06451727+0023454      &    F2V   &  $-16.58\pm0.30$   &   0.31$\pm$0.06 & 2.30$\pm$0.20  & 2.69  &   3.831   &  1.00    & 1.6  &  6.0  & 1, 3, 6     \\ 
J06451841+0022189       &    G4V   &  $-5.20\pm0.30$    &   2.05$\pm$0.10 & 2.70$\pm$0.05  & 1.05  &   3.750   &  0.19    & 1.0  &  10.0  & 6		  \\
 J06452454+0022448      &    G8V   &  $-21.18\pm0.25$   &   2.20$\pm$0.30 &   -            & 2.33  &   3.727   &  0.75    & 1.8  &  1.5  & 6		\\
D25 J06452476+0013360   &     -    &  $-30.50\pm0.30$   &    -	          &    -	   &   -   & -         &      &   -	   &  -    & 4  	\\ 
J06452774+0015514       &    K4V   &  $-29.00\pm1.30$   &      -          &  -             & 2.77  &   3.638   &   0.06   & 0.7  &  1.5  & 2		  \\
J06451118+0011402   &      M1V     &   $-3.88\pm0.12$   &     noise       &  7.40$\pm$0.34 & 0.17    & 3.566   &   -0.42  & 0.3 & 2.0&  Potential PMS, 4 \\
J06453567+0018473    &      K7V     &   $-3.20\pm0.05$  &   2.79$\pm$0.10 &  6.00$\pm$0.15 & 0.21    & 3.610   &   -0.20  & 0.4 & 2.0 &  Potential PMS, 6	\\ 
& & & & & & & & & &\\ 
\hline
\hline	
\end{tabular}
\begin{flushleft}
$*$For this star we obtained a spectrum with the BFOSC instrument at the 1.5~m Tel. at the Loiano Observatory (http://www.bo.astro.it/loiano/index.htm),\\
 using grism $\#4$ with an exposure of 20~minutes.\\
$^1$ PMS $\delta$ Scuti  candidates. \\
$^2$ Object matching Class I source in P09.\\
$^3$ Object matching Class II source in P09.\\
$^4$ Star member of Dolidze 25. \\
$^5$ Star member of Cl2 (see definition in P09). \\
$^6$ Star member of RN (see definition in P09). \\
$^7$ Star member of RW (see definition in P09). \\
\end{flushleft}
\vspace{1cm}
\caption{Physical parameters of Field stars with emission in the H$\alpha$ line. Columns are the same as in Table 3.}
\label{tab4} 
\begin{tabular}{c c c c c  c c c c c l}
\hline\hline 
id   & SpT & EW($\rm{H_\alpha}$)   & EW(Ca I+ Fe I ) &  EW(Na I) & Av   & $\log \rm{T_\mathrm{eff}}$  & $\log \rm{L_{BC}/L_\odot}$ & Mass &age   &  Notes \\
     &     &   [\AA]               &  [\AA]          & [\AA]     &(mag) &                             &                            & (M$_\odot$)  & (Myr)    & \\
\hline 
\hline 
 & & & & & & & & & &\\
 J06445227+0014300   &      M2.5V   &   $-5.10\pm0.23$      &     2.39$\pm$0.13  &  10.00$\pm$0.67   & 0.00  & 3.519  & -& - & - &  Dme  \\
J06445433+0021190    &      M4.5V   &   $-20.71\pm0.34$     &     1.22$\pm$0.40  &  3.20$\pm$0.34    & 0.36  & 3.482  & -& - & - &  Dme \\
J06450151+0008493    &      M2.5V   &   $-7.20\pm0.10$      &     1.15$\pm$0.30  &  5.25$\pm$0.20    & 0.00  & 3.547  & -& - & - &  Dme \\ 
D25 06452246+0015408 &      M4V     &   $-8.22\pm0.23$      &     noise	         &    noise 	     & 0.00  & 3.510  & -& - & - &  Dme     \\ 
J06453429+0022261    &      M2V     &   $-1.58\pm0.23$      &     2.37$\pm$0.21  &  8.79$\pm$0.23    & 0.00  & 3.510  & -& - & - &  Dme\\
 J06453550+0024158   &      M2V     &   $-11.90\pm0.25$     &      -	         &  9.00$\pm$0.20    & 1.29  & 3.550  & -& - & - &  Dme\\
  J06453643+0011254  &     M3.5V    &   $-3.90\pm0.23$      &     2.54$\pm$0.20  &  8.95$\pm$0.40    & 0.00  & 3.506  & -& - & - &  Dme \\
D25 J06454297+0011461&      M4V     &   $-9.15\pm0.15$      &     3.89$\pm$0.40  &  4.01$\pm$0.80    & 0.00  & 3.510  & -& - & - &  Dme   \\
J06454948+0011347    &      M4V     &   $-0.18\pm0.10$      &      -             &    -              & 0.00  & 3.493  & -& - & - &  Dme \\
 J06455486+0008134   &      M2V     &   $-8.50\pm0.20$      &      -	         &  2.89$\pm$0.20    & 0.52  & 3.550  & -& - & - &  Dme  \\
 & & & & & & & & & &\\
\hline
\hline	
\end{tabular}
\end{table*}

\section*{Acknowledgments}
We are greatful to the referee Prof. W. Lawson for his useful comments
  and suggestions.
This work was supported by the Italian ESS project, contract ASI/INAF
I/015/07/0, WP 03170 and by the European Helio- and Asteroseismology
Network (HELAS), a major international collaboration funded by the European
Commission's Sixth Framework Programme. EP is funded by the Spanish MICINN under the Consolider-Ingenio 2010
Program grant: First Science with the GTC.
This research has made use of the SIMBAD database, operated at CDS,
Strasbourg,
France. This publication makes use of data products from the Two Micron
All Sky
Survey, which is a joint project of the University of Massachusetts and the
Infrared Processing and Analysis Center/California Institute of
Technology, funded
by the National Aeronautics and Space Administration and the National
Science
Foundation.

\label{lastpage}

\end{document}